\documentclass[11pt,a4paper]{article}

\usepackage{amsmath,amsthm,amssymb,amsfonts}

\usepackage{authblk}
\usepackage{setspace}
\usepackage{mathrsfs}
\usepackage{cases}
\usepackage{graphicx}
\usepackage{enumerate}
\usepackage{subcaption} 
\usepackage{datetime}

\numberwithin{equation}{section}
\newcommand{\ii}{{\rm i}}
\newcommand{\ee}{{\rm e}}
\newcommand{\dd}{{\rm d}}
\newcommand{\uu}{\bar{u}}
\newcommand{\x}{{\rm x}}
\newcommand{\y}{{\rm y}}

\newcommand{\I}{1\!\!1}
\newcommand{\vol}{{\rm vol}}

\theoremstyle{definition}

\usepackage{color}
\newdateformat{daymonthyear}{\THEDAY \, \monthname[\THEMONTH] \THEYEAR}
\newdateformat{monthyear}{\monthname[\THEMONTH] \THEYEAR}

\begin{document}

\title{Quantum fields during black hole formation: \\ How good an approximation is the Unruh state?}
\author{Benito A. Ju\'arez-Aubry$^{1}$\thanks{{\tt benito.juarez@correo.nucleares.unam.mx}} }
\author{Jorma Louko$^{2}$\thanks{\tt jorma.louko@nottingham.ac.uk}}
\affil{$^{1}$Departamento de Gravitaci\'on y Teor\'ia de Campos \\Instituto  de  Ciencias  Nucleares,  Universidad  Nacional Aut\'onoma de M\'exico, \\A. Postal 70-543, Mexico City 045010, Mexico}
\affil{$^{2}$School of Mathematical Sciences, University of Nottingham, Nottingham NG7 2RD, UK}
\date{\daymonthyear\today}

\maketitle

%\centerline{\textbf{Not for circulation.}}

\begin{abstract}
We study the quantum effects of a test Klein-Gordon field in a Vaidya spacetime consisting of a collapsing null shell that forms a Schwazschild black hole, by explicitly obtaining, in a $(1+1)$-dimensional model, the Wightman function, the renormalised stress-energy tensor, and by analysing particle detector rates along stationary orbits in the exterior black hole region, and make a comparison with the folklore that the Unruh state is the state that emerges from black hole formation. In the causal future of the shell, we find a negative ingoing flux at the horizon that agrees precisely with the Unruh state calculation, and is the source of black hole radiation, while in the future null infinity we find that the radiation flux output in the Unruh state is an upper bound for the positive outgoing flux in the collapsing null shell spacetime. This indicates that back-reaction estimates based on Unruh state calculations over-estimate the energy output carried by so-called pre-Hawking radiation. The value of the output predicted by the Unruh state is however approached exponentially fast. Finally, we find that at late times, stationary observers in the exterior black hole region in the collapsing shell spacetime detect the local Hawking temperature, which is also well characterised by the Unruh state, coming from right-movers. Early-time discrepancies between the detector rates for the Unruh state and for the state in the collapsing shell spacetime are explored numerically.
\end{abstract}

\singlespacing

%=====================================================================================================
% INTRODUCTION
%=====================================================================================================
\section{Introduction}
\label{sec:Intro}

Quantum field theory in curved spacetimes has made several remarkable physical predictions in the past years. A particularly notable one is that of black hole thermal radiation, predicted by Hawking \cite{Hawking:1974sw}, which eventually lead to the discovery of black hole evaporation and gave birth to the black hole information loss puzzle \cite{Hawking:1976ra}, which remains open. Since then, the study of black hole radiation in field theory has become a large industry of theoretical and mathematical physics. 

There exists the common folklore that the modelling of the state of a test quantum field (typically Klein-Gordon) in a collapsing star spacetime, which eventually forms a black hole at late times, is well described by considering the analogous system of a quantum field propagating in Schwarzschild spacetime in the Unruh state, a stationary but time-reversal non-invariant state introduced by Unruh in \cite{Unruh:1976db}. In particular, computations with the Unruh state show that there is a negative energy flux across the Schwarzschild horizon that is the source of Hawking radiation.

It is clear, however, that this model can only be good (extremely good, as we shall see in this work) in certain spacetime regions, since stellar collapse is a non-stationary process, and hence, the early time quantum effects are not captured by this analogy. This observation has produced some speculation that early-time processes may avoid the formation of black holes altogether. Recently, the works \cite{Kawai:2013mda, Kawai:2014afa, Kawai:2015uya, Kawai:2017txu, Baccetti:2016lsb, Baccetti:2016dzy, Baccetti:2017ioi, Baccetti:2017oas} are in this vein. On the other hand, the recent works \cite{Arderucio-Costa:2017etb, Chen:2017pkl, Unruh:2018jlu} argue in the oposite direction, namely, that early time quantum processes -- pre-Hawking radiation -- cannot prevent the horizon formation, based on simple models of stellar collapse. It is also worth reminding ourselves of earlier work in the same spirit of \cite{Arderucio-Costa:2017etb, Chen:2017pkl} that is based on the Unruh state analysis \cite{Paranjape:2009ib}, as well as \cite{Davies:1976ei}, which developed {\it en passant} powerful conformal techniques for $(1+1)$-dimensional conformally-coupled theories, and pointed at the connection between Hawking radiation and the conformal anomaly of the renormalised stress-energy tensor, established in \cite{Christensen:1977jc}.

In the present paper, we take on the question of how good the Unruh state is in modelling quantum effects during black hole formation. Our framework is the following: we consider a spherically symmetric collapsing null shell spacetime, such that the exterior region of the shell is isometric to Schwarzschild spacetime. We then consider a test quantum Klein-Gordon field and construct a natural state in the collapsing shell spacetime, and compare the quantum effects in this state with those of the Unruh state in Shwarzschild. We do this exactly in $1+1$ dimensions by considering incoming modes of positive frequency with respect to $\partial_v$ at $\mathscr{I}^-$ and imposing Dirichlet boundary conditions at the origin of the radial coordinate, $r = 0$, to emulate a $(3+1)$-dimensional spherically symmetric spacetime. This allows one to construct the Wightman function explicitly by standard sum-over-modes techniques, much like is done in receding mirror spacetimes \cite{Davies:1976hi, Davies:1977yv}, which have also been very useful models in characterising the emergence of thermal radiation in non-stationary situations. See e.g. the recent work \cite{Good:2016oey, Good:2017kjr, Good:2017ddq}, as well as \cite{Juarez-Aubry:2014jba} for the detector responses in the context of receding mirrors.

We give a closed-form expression for the Wightman function in the $1+1$ collapsing null shell spacetime without any approximations, and obtain both the renormalised stress-tensor in this state by conformal techniques \cite{Davies:1976ei} and the detector rates of late-time observers with a derivative coupling to the field \cite{Juarez-Aubry:2014jba}, so as to measure the experience of local observables. The derivative coupling is chosen in order to avoid the problems stemming from infrared ambiguities in the Wightman function that are well-known to appear in $1+1$ dimensions for conformal theories, but which do not plague the renormalised stress-enegy tensor.\footnote{Incidentally, the two-point function that we construct is absent of these ambiguities, but since our purposes are to compare our findings with the Unruh state, we systematically adhere to the derivative-coupling Unruh-DeWitt detector.}

We then go on to compare these quantum effects produced by the field in the collapsing shell spacetime state with those produced in the Unruh state. The comparison can be readily made in the region to the future of the shell, where the spacetime is isometric to Schwarzschild and one can therefore identify the two theories in a precise way, as a field algebra equipped with either the state that we construct in this work or the Unruh state. 

In the case of the renormalised stress-energy tensor, we find that the behaviour near the horizon is matched very precisely by the Unruh state, to order $O\left( (r-2M)^2 \right)$ and is indeed dominated by a negative ingoing flux. Near future null infinity, in both cases there is a positive outgoing flux of energy carrying Hawking radiation, but in the case of the collapsing shell there is a retarded time dependence as $r \to \infty$ (or the advanced time $v \to \infty$), with this dependence being a strictly non-decreasing function of the $u$-time. This flux at future null infinity however approaches exponentially fast (in $u$-time) the value of the flux predicted using the Unruh state. The values for the two stress-energy tensor coincide at future timelike infinity, as discussed below eq. \eqref{TuuShellInfty}.  A numerical comparison of the two states is plotted, showing the early-time discrepancies in the two cases.

In the case of local observers equipped with particle detectors, we see that a detector that is switched on at the shell crossing\footnote{The switch-on time is however irrelevant near future null infinity, and this is but a particular illustrative choice.} and travels towards future timelike inifinity at fixed radial coordinate will detect at late times Hawking radiation at the local Hawking temperature. This is also known to be the case in the Unruh state, and we therefore see that late time observers register the thermal character of the black hole radiation in this more realistic setting. The early-time discrepancies in the rates are analysed by numerical techniques.

We shall therefore conclude that the Unruh state is an excellent approximation in the near horizon regime description of the quantum processes during black hole formation. In particular, it captures in an excellent way the negative energy flux across the horizon giving rise to black hole radiation. Moreover, when analysing the radiation output at future null infinity in the exterior black hole region ($v \to \infty$) of the collapsing shell spacetime, it can be verified that the outgoing flux (see eq. \eqref{TuuShellInfty} below) is strictly non-decreasing in $u$-time, showing that the largest output of radiation comes from the near-horizon region. Thus, making it unlikely that pre-Hawking radiation can account for black hole formation avoidance. Indeed, the stress-energy flux output at any point along $\mathscr{I}^+$ in the collapsing shell scenario is bounded by the constant outgoing flux calculated from the Unruh state (see eq. \eqref{TuuUnruhInfty} below). Thus, any back-reaction evaluation based on Unruh state estimates \cite{Paranjape:2009ib} (and also on the one in \cite{Davies:1976ei}) is already over-estimating the early-time outgoing radiation flux at infinity.

This paper is organised in the following way: In Section \ref{sec:KG} we provide the geometric and quantum-field-theoretic preliminaries of the Klein-Gordon quantum theory in Schwarzschild spacetime (and its maximal extension), stressing the r\^ole played by the spacetimes isometries in the definition of states, while at the same time introducing some of the notation for this paper. In Section \ref{sec:Null} we describe the geometry of the collapsing null shell spacetime and in $1+1$ dimensions obtain the state of a Klein-Gordon field propagating on this spacetime, which is defined in terms of positive frequency modes at $\mathscr{I}^-$ with respect to $\partial_v$ and  with Dirichlet boundary conditions at $r = 0$, and which emulates the state of a Klein-Gordon field in a $(3+1)$-dimensional, spherically-symmetric, collapsing null shell spacetime. We then obtain the stress-energy tensor of the Klein-Gordon field in this state in Section \ref{sec:Stress-Energy}, and compare it with the Unruh-state stress-energy in Schwarzschild. The comparison can be done unambiguously in the region in the causal future of the shell because this region is isometric to part of the maximally extended Schwarzschild spacetime. In this section, we also provide a comparative analysis for the near-horizon and near-future-null-infinity regimes for the two states, finding that the ingoing negative energy flux at the horizon, characteristic of the Unruh state, is in excellent agreement with the collapsing null shell scenario. We also provide the analysis for the flux radiation output at $\mathscr{I}^+$, as discussed above. In Section \ref{sec:Detectors} we show that the black hole radiation in the collapsing shell spacetime is of thermal character at late times, by analysing the rate of a sharply-switched detector coupled to the derivative of the Klein-Gordon field. We find that near $i^+$ the rate registered from the right-movers by a detector following an orbit generated by $\partial_t$ in the exterior black hole region to the future of the shell, which is an orbit at fixed radial coordinate, is thermal at the local Hawking temperature, i.e., the detector registers the Hawking temperature weighted by an appropriate Tolman factor. The rate near future timelike infinity is in excellent agreement with the rate detected from a coupling to a field in the Unruh state. Finally, our conclusions appear in Section \ref{sec:conc}.

Throughout this paper, by a spacetime, $(M,g)$, we mean a real $n$-dimensional, connected (Hausdorff, paracompact) differentiable manifold, $M$, equipped with a Lorentzian metric $g$ with signature $(-, +, \ldots, +)$. Our spacetimes of interest are additionally time-orientable and globally hyperbolic \cite{BernalSanchez1, BernalSanchez2}. We use units in which the speed of light, the reduced Planck's constant and Newton's constant have value unit, $c = \hbar = G_\text{N} = 1$, and we further fix Boltzmann's constant as $k_{\rm B} = 1$. Spacetime points are denoted by Roman characters ($\x, \y, \dots$). Abstract tensor indeces are denoted by latin characters, $a, b, \ldots$. Complex conjugation is denoted by an overline. %Concrete operators on Hilbert spaces are indicated by capital letters surmounted with carets, such as $\widehat{A}, \widehat{B}, \dots$, while elements of an abstract non-commutative algebra are indicated by caret-free letters such as $A, B, \ldots$. 
The adjoint of a Hilbert-space operator, $\widehat A$, is denoted by ${\widehat A}^*$.  $O(x)$ denotes a quantity for which $O(x)/x$ is bounded as $x \to 0$ and $o(x)$ is such that $o(x)/x \to 0$ in the limit under consideration.

%=====================================================================================================
% UNRUH
%=====================================================================================================
\section{The Klein-Gordon field in Schwarzschild spacetime}
\label{sec:KG}

In this section, we briefly recall some elements of quantum field theory in Schwarzschild spacetime. This will also serve the purpose of introducing the relevant notation for this work. First, we shall recall the geometric structure of Schwarzschild spacetime and its maximal extension, reminding ourselves the large symmetry structure of these spacetimes, which serves as a guideline for constructing the quantum theory states. We shall then introduce a Klein-Gordon field in the Schwarzschild maximal extension and recall the properties of the usual states of the theory, the Boulware, Unruh and Hartle-Hawking-Israel states, emphasising the physical relevance of the Unruh state.

\subsection{Geometric preliminaries}

For the purposes of the discussion of field theory global states in Schwarzschild spacetime, it is useful to consider the maximal Kruskal-Szekeres extension, to which we refer as Kruskal spacetime, and to consider states on field algebras defined thereon. Kruskal spacetime, $(M_{\rm K}, g_{\rm K})$, is defined by the underlying manifold $M_{\rm K} = \mathbb{R}^2 \times \mathbb{S}^2$ equipped with the metric
\begin{equation}
g_{\rm K} = -\frac{32 M^3 \ee^{-r/(2M)}}{r} \dd U \dd V + r^2\left( \dd \theta^2 + \sin^2 \theta \dd \phi^2 \right)
\end{equation}
where $U \in \mathbb{R}$, $V \in \mathbb{R}$, $\theta \in [0, \pi]$ and $\phi \in [0, 2 \pi)$ are global coordinates, $r$ is a non-negative global spacetime function defined by $[r(U,V)/(2M)-1]\exp[r(U,V)/(2M)] = -UV$ and $M>0$ is a length parameter corresponding to the ADM black hole mass. A spacetime singularity is located at $r = 0$, and the spacetime is asymptotically flat as $r \to \infty$. The spacetime is globally hyperbolic; for example $U + V = 0$ is a Cauchy surface.

The isometry group of Kruskal spacetime is generated by the spacelike Killing vector fields that generate the spherical symmetry, $\zeta_1 = \partial_\phi $, $\zeta_2 = \sin \phi \partial_\theta + \cot \theta \cos \phi \partial_\phi$ and $\zeta_3 = \cos \phi \partial_\theta - \cot \theta \sin \phi \partial_\phi$, together with $\xi = (4M)^{-1} (-U \partial_U + V \partial_V)$. The Killing vector $\xi$ defines a bifurcate Killing horizon (by $\xi^a \xi_a = -(1-2M/r) = 0$), which is located at $r = 2M$ and separates the spacetime into four regions, covered by the charts indicated in Table \ref{Table:Kruskal}. The Killing horizon may be decomposed as ${\rm H} = {\rm H}^+ \cup {\rm H}^-$, where ${\rm H}^-$ is located at $V = 0$ and ${\rm H}^+$ at $U = 0$. %See Fig. \ref{Fig:Kruskal} for the conformal diagram of Kruskal spacetime.

\begin{table}
\begin{center}
  \begin{tabular}{ l  l c c  c }
    \hline
     & Region & sgn$(\xi^a \xi_a)$ &sgn$(U)$ & sgn$(V)$ \\ \hline
    I: Exterior & $2M < r$ & $-1$ & $-1$ & $+1$ \\ %\hline
    II: Black hole & $0 < r < 2M$ & $+1$ & $+1$ & $+1$ \\
    III: White hole & $0 < r < 2M$ & $+1$ & $-1$ & $-1$ \\
    IV: Isometric exterior & $2M < r$ & $-1$ & $+1$ & $-1$ \\
    \hline
  \end{tabular}
  \caption {Regions of the Kruskal-Szekeres extension of the Schwarzschild black hole.}
  \label{Table:Kruskal}
\end{center}
\end{table}

%\begin{figure}
 % \centering
  %\includegraphics[width=0.5\linewidth]{Kruskal.pdf}
  %\caption{Conformal diagram of Kruskal spacetime.}
  %\label{Fig:Kruskal}
%\end{figure}

An interesting region of the maximally extended Schwarzschild spacetime is the region $V>0$, which consists of Regions I and II and the portion of the Killing horizon joining them. In this region, we can introduce an ingoing Eddington-Finkelstein coordinate with the transformation $V = \exp[v/(4M)]$, $v \in \mathbb{R}$, and view this submanifold as an asymptotically flat, globally-hyperbolic ``ingoing Eddington-Finkelstein" spacetime,\footnote{Beware that Cauchy surfaces in Kruskal spacetime do not necessarily restrict to Cauchy surfaces in the non-maximally extended Schwarzschild spacetime.} $(M_{\rm S}, g_{\rm S})$ as the underlying manifold $M_{\rm S} = \mathbb{R}^2 \times \mathbb{S}^2$ equipped with the metric induced from Kruskal spacetime,
\begin{equation}
g_{\rm S} = -\frac{8 M^2 \ee^{-r/(2M)+v/(4M)}}{r} \dd U \dd v + r^2\left( \dd \theta^2 + \sin^2 \theta \dd \phi^2 \right).
\label{gSchw}
\end{equation} 

The group of isometries in this region is inherited from Kruskal, and generated by the Killing vectors $\zeta_1, \zeta_2$ and $\zeta_3$, together with the restriction of $\xi$, which can be written globally as $\xi = -(4M)^{-1} U \partial_U + \partial_v$.  

The exterior region of Schwarzschild, Region I with $U < 0$ and $V >0$, is covered by the familiar Schwarzschild coordinates. They are related to the global coordinates by introducing for $U < 0$ the outgoing Eddington-Finkelstein coordinate $u = -4M \ln (-U)$, and further relating $v = t + r_*$, $u = t - r_*$, where $r_* = r + 2M \ln [r/(2M) - 1]$ is the tortoise radial coordinate. The Schwarzschild metric acquires the familiar form
\begin{equation}
g_{\rm S} = - \left( 1 - \frac{2M}{r} \right) \dd t^2 + \left( 1 - \frac{2M}{r} \right)^{-1} \dd r^2 + r^2\left( \dd \theta^2 + \sin^2 \theta \dd \phi^2 \right).
\end{equation}

The isometry group of the exterior region is generated by $\zeta_1, \zeta_2$ and $\zeta_3$, together with the restriction of $\xi$, which can be written as $\xi = \partial_t$.

\subsection{States in (the maximal extension of) Schwarzschild spacetime}

It is useful to think of the real Klein-Gordon field in an arbitrary globally-hyperbolic curved spacetime, $(M,g)$, in terms of smeared fields, $\Phi(f)$ for test functions $f \in C_0^\infty(M)$,\footnote{Formally, we can represent $\Phi(f) = \int_{M} \! \dd \vol(\x) \Phi(\x) f(\x)$.} generating an abstract $^*$-algebra with identity $\I$, $\mathscr{A}(M)$, satisfying the following axioms: 

Let $f, g \in C_0^\infty(M)$, then (i) $f \mapsto \Phi(f)$ is linear (linearity), (ii) $\Phi(f)^* = \Phi(f)$ (hermiticity), (iii) $\Phi((\Box - m^2 - \xi R)f) = 0$ (field equation) and (iv) $[\Phi(f),\Phi(g)] = - \ii E(f,g)$ (spacetime commutation relations). Here $E = E^- - E^+$ is the advanced-minus-retarded classical causal propagator of the Klein-Gordon equation (see. e.g. \cite{Benini:2013fia}), where $E^-$ and $E^+$ are the advanced and retarded Green operators of the Klein-Gordon equation, that can be regarded as a bi-distribution taking $f, g \in C_0^\infty(M)$ to 
\begin{equation}
E(f,g) = \int_{M \times M} \! \dd \vol(\x) \dd \vol(\x') f(\x) E(\x, \x') g(\x').
\end{equation}

It is guaranteed to exist and be unique because the Klein-Gordon operator is normally hyperbolic (having unique $E^\mp$), and satisfies $(\Box - m^2 - \xi R) E f = 0$, i.e., $Ef(\x) = \int_{M} \! \dd \vol(\x') E(\x, \x') f(\x')$ solves the classical Klein-Gordon equation. In fact, all smooth solutions to the classical Klein-Gordon equation with initial data of compact support are of the form $\phi_f = Ef \in {\rm Sol}_{\rm KG}$, with $f \in C_0^\infty(M)$. See Lemma 3.2.1 in \cite{Wald:1995yp} for the argument in Minkowski space that can be extended to globally hyperbolic spacetimes.

We refer to $\mathscr{A}(M)$ as the {\it real Klein-Gordon algebra} and, from this point of view, the Klein-Gordon field is an algebra-valued distribution.\footnote{In $1+1$ dimensions, it is more convenient to think of the algebra of {\it derived fields}.}

{\it States} are linear functionals $\omega: \mathscr{A}(M) \to \mathbb{C}$, such that they are (i) normalised, $\omega(\I) = 1$ and (ii) positive, for $A \in \mathscr{A}(M)$, $\omega(A A^*) \geq 0$, and they are determined by the specification of all the $n$-point functions of the form $\omega(\Phi(f_1) \ldots \Phi(f_n))$. Of particular relevance are the {\it quasi-free} or Gaussian states, which are determined fully by the two-point function, via the relation $\omega(\exp[\ii \Phi(f)]) = \exp[-\omega(\Phi(f)\Phi(f))/2]$. Vacuum states are, in particular, quasi-free.

The standard textbook approach, where fields are operator-valued distributions acting on a Hilbert space, can be recovered using the GNS construction. Out of the Klein-Gordon algebra, $\mathscr{A}(M)$, and a state, $\omega$, on the algebra, there is a standard procedure to construct a {\it GNS triple} $(\pi, \mathscr{D} \subset \mathscr{H}, \Omega)$, where $\pi: \mathscr{A}(M) \to \mathscr{L}(\mathscr{D})$ is a representation with respect to the state $\omega$  that maps elements of the algebra to operators on a dense subspace $\mathscr{D} \subset \mathscr{H}$ of the Hilbert space $\mathscr{H}$ and where $\Omega \in \mathscr{H}$ is a cyclic vector, which means that span$\{\pi(A) \Omega\}$ (with $A \in \mathscr{A}$) is dense in $\mathscr{H}$, that we identify with the vacuum.  The two-point function is $\omega(\Phi(f)\Phi(g)) = \langle \Omega | \pi(\Phi(f)) \pi(\Phi(g)) \Omega \rangle$. See \cite[Chap. 5.1.3]{Khavkine:2014mta} for an overview of the GNS construction.

We denote $\widehat{\Phi} = \pi(\Phi)$. In terms of the {\it one-particle structure of the Hilbert space with respect to} $\omega$, $(K, \mathcal{H})$, which is such that $\mathscr{H}$ is the symmetric Fock space built out of the one-particle Hilbert space $\mathcal{H}$, $\mathscr{H} = \oplus_{n = 0}^\infty \mathcal{H}^{\circ n}$, with a prescribed notion of positive frequency given by the polarisation map $K: {\rm Sol}_{\rm KG} \to \mathcal{H}$ (mapping classical Klein-Gordon solutions to positive-frequency Hilbert space vectors) \cite[Chap. 2.3]{Wald:1995yp}, the operator $\widehat{\Phi}$ can be written in terms of annihilation and creation operators acting on Fock space, respectively $\widehat{a}: \mathscr{H} \to \mathscr{H}$ and $\widehat{a}^*: \mathscr{H} \to \mathscr{H}$ as $\widehat{\Phi}(f) = \ii \widehat{a}(K E f) - \ii \widehat{a}^*(K E f)$. The details can be found in \cite{Wald:1995yp, Kay:1988mu}.

For the problem at hand, that of field theory in Schwarzschild spacetime, one can construct three distinguished states, that are invariant under either (i) the isometries of the whole Kruskal-Szekeres extension of Schwarzschild spacetime, $(M_{\rm K}, g_{\rm K})$, (ii) the isometries of the non-maximally extetended Schwarzschild black hole, comprising regions I and II and the portion of the horizon joining them, $(M_{\rm S}, g_{\rm S})$ or (iii) the isometries of the exterior region of Schwarzschild, correspoding to Region I in the Kruskal extension.

Case (i) is known as the Hartle-Hawking-Israel state, defined in the whole Kruskal manifold, where it can be seen as a map $\omega_{\rm HHI}: \mathscr{A}(M_{\rm K}) \to \mathbb{C}$. It was conjectured to exist initially in \cite{Hartle:1976tp, Israel:1976ur}, shown to be unique in \cite{Kay:1988mu} and constructed in four spacetime dimensions in \cite{Sanders:2013vza}. It has modes of positive and negative frequency with respect to the generators of the Killing horizon, $\partial_U$ and $\partial_V$. It represents the state for an eternal black hole in equilibrium. Further, the restriction of the state to Region I is a KMS (thermal equilibrium) state at the Hawking temperature. Case (iii) is known as the Boulware state. It is defined in the exterior region of Schwarzschild and was initially studied in \cite{Boulware:1974dm}. It has modes of positive and negative frequency with respect to the exterior Schwazschild timelike Killing vector field $\xi = \partial_t$, and fails to be regular at the event horizon.

The relevant setting for us is case (ii), the Unruh state, which is defined as a map in the ``ingoing Eddington-Finkelstein" region of the Kruskal space, and hence can be thought of as a map $\omega_{\rm U}: \mathscr{A}(M_{\rm S}) \to \mathbb{C}$, whose state vector we denote by $|\Omega_U \rangle$. This state was introduced in \cite{Unruh:1976db} to mimick the late-time quantum behaviour on a black hole produced by stellar collapse, and first abstractly constructed in $3+1$ dimensions in \cite{Dappiaggi:2009fx}. It is obtained by prescribing modes of positive frequency on the Cauchy surface $\Sigma = \mathscr{I}^- \cup H^-$, obtained from the union of the past null infinity, $\mathscr{I}^-$, located at $U \to -\infty$, with the past event horizon, located at $V = 0$ in the Kruskal spacetime (corresponding to $v \to - \infty$). The positive frequency on the past horizon is prescribed with respect to the horizon generator, $\partial_U$ (with $U$ being the affine parameter along $H^-$), whereas the positive frequency on past null infinity is prescribed with respect to the vector field $\partial_v$, the null geodesic generator of $\mathscr{I}^-$ (with $v$ being the affine parameter along $\mathscr{I}^-$). 

Moreover, since its introduction, the Unruh state has been a key ingredient in the study of black hole radiation, and its eventual evaporation. In particular, it is this state that is considered for obtaining the Hawking temperature that can be recorded at $i^+$ by an observer, as well as the flux of stress-energy carried away by Hawking radiation. See for example \cite{Paranjape:2009ib}, in particular Appendix A therein, for a discussion on the choice of the Unruh state in this context. As mentioned in the Introduction, the purpose of this work is to study in a simple $(1+1)$-dimensional model how the qualitative features captured by the Unruh state compare with the actual properties a state in a spacetime of stellar collapse that is non-stationary in the exterior region. The reasons to favour a $(1+1)$-dimensional treatment are, first, that it allows for explicit computations with analytic control and, second, that they provide a good estimate on the amount of radiation that can escape to infinity in $3+1$ dimensions, as argued in \cite{Ring:2006fk}, and also discussed in \cite{Arderucio-Costa:2017etb}.

A remark is due at this stage.  While it is true that the Boulware and HHI state Wightman functions in $1+1$ dimensions are invariant under the isometries generated by the Killing vector $\xi$, the Unruh state Wightman function changes under these isometries by an additive constant \cite{Fulling-Ruijsenaars, Kay:2000fi}. The origin of this issue is the well-known infrared ambiguity of the $(1+1)$-dimensional conformally coupled Klein-Gordon field \cite{Dappiaggi:2009fx}. Nevertheless, the Unruh state may still be regarded as invariant under the isometries in the sense that the stress-energy tensor and other quantities built from derivatives of the Wightman function are invariant. In higher dimensions this situation does not arise, and the Unruh state is invariant in the standard sense.

%=====================================================================================================
% COLLAPSING SHELL
%=====================================================================================================
\section{The Klein-Gordon field in a collapsing null shell spacetime}
\label{sec:Null}

In this section, we study the Klein-Gordon theory in an ingoing Vaidya spacetime \cite{Vaidya:1951zz} with a discontinuous length function, ${\rm M}$, which represents a thin, ingoing, spherically-symmetric light pulse, a null shell, with total energy $M$, that forms a black hole. We show how to construct a state for the quantum Klein-Gordon field in $1+1$ dimensions by writing two-point function. Importantly, in the exterior region of the black hole produced by the collapsing shell, to the future of the shell, the state that we construct is not invariant with respect to the isometry of Schwazschild spacetime, generated by $\xi = \partial_t$, unlike the Unruh state (up to an additive constant). As a consequence, an observer moving at a fixed radius will detect, by reading his trust-worthy particle detector, time-dependent radiation coming from the black hole, and the flux of Hawking radiation will also be time-dependent.

\subsection{Geometry of the collapsing null shell spacetime}

The spacetime that we consider, which we denote by $(M_{\rm V}, g_{\rm V})$ has as an underlying manifold $\mathbb{R}^2 \times \mathbb{S}^2$, and it is equipped with the metric
\begin{subequations}
\begin{align}
& g_{\rm V} = -\left(1 - \frac{2\mathrm{M}(v)}{r} \right) \dd v^2 + 2 \dd v \dd r + r^2 \left( \dd \theta^2 + \sin^2 \theta \dd \phi^2 \right), \\
& \mathrm{M}(v)  =\left\{
                \begin{array}{ll}
                  0, &\text{ if } v < 0, \\
                  M, &\text{ if } v \geq 0,
                \end{array}
              \right. 
\end{align} 
\end{subequations} 
where $v \in \mathbb{R}$ is a null coordinate, $r \in (0, \infty)$ is a radial coordinate and $\theta \in [0, \pi]$ and $\phi \in [0, 2 \pi)$ are angular coordinates. The shell is located at $v = 0$, along a null surface $S$. For $v \geq 0$ the spacetime is isometric to a portion of Schwazschild spacetime with length parameter $M$, with $v$ playing the r\^ole of the advanced Eddington-Finkelstein coordinate, and a standard change of coordinate puts the metric in the form of eq. \eqref{gSchw} in this region. For $v < 0$ the spacetime is isometric to a portion of Minkowski spacetime, with $v$ playing the r\^ole of a Minkowski outgoing light-cone coordinate. We denote the black hole horizon by H$^+$. The shell falls into the singularity at $(v,r) = (0,0)$ or, in terms of Kruskal coordinates that cover the causal future of the shell, at $(U,V) = (1,1)$.

\begin{figure}
  \centering
  \includegraphics[width=0.65\linewidth]{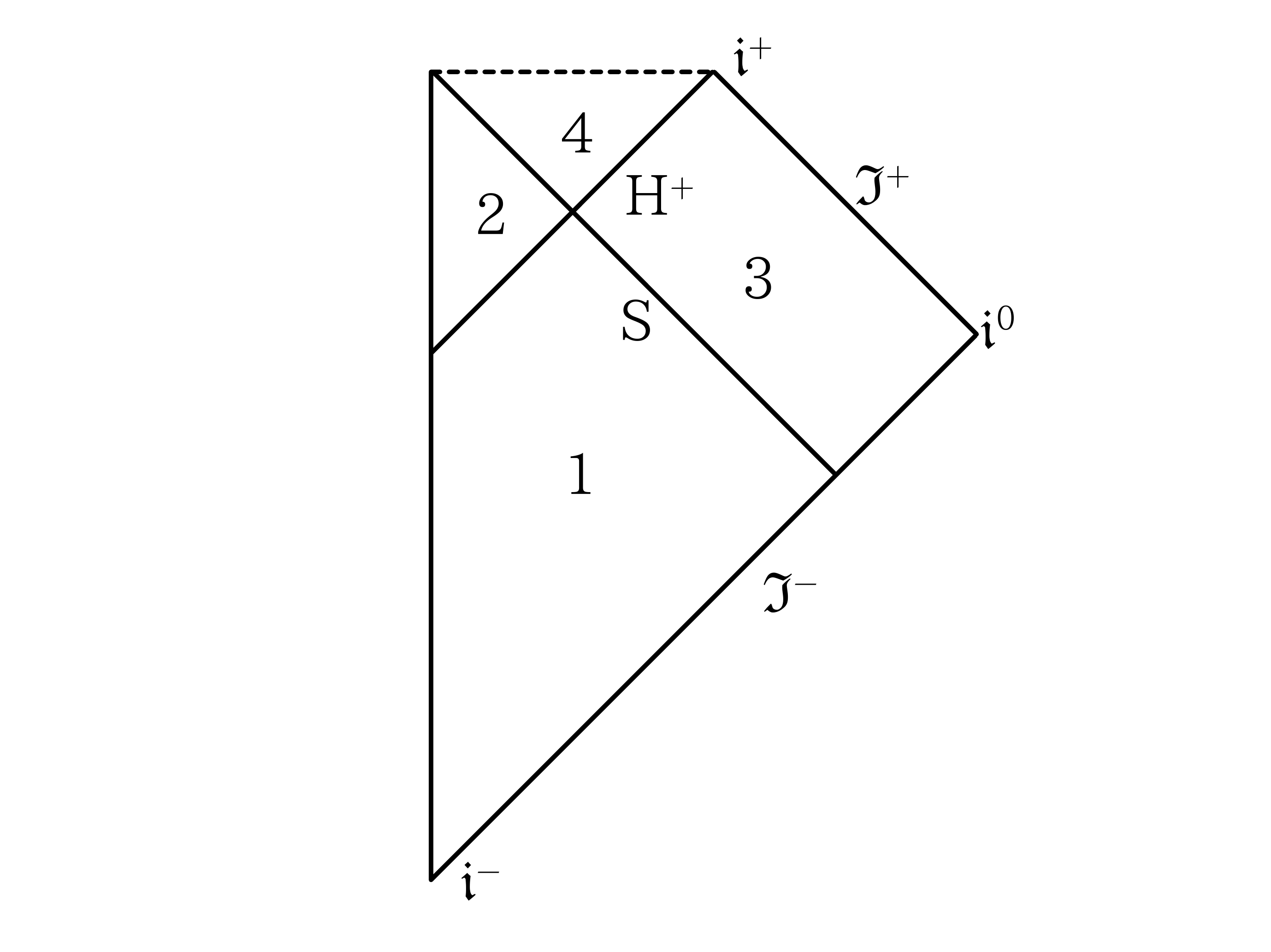}
  \caption{Conformal diagram of the Vaidya spacetime describing a collapsing null shell that forms a black hole.}
  \label{Fig:Shell}
\end{figure}

The spacetime can be thought of as the union of 4 regions, as shown in the conformal diagram in Fig. \ref{Fig:Shell}. We call the exterior black hole region to the past of the shell, not including the shell, $J^-(S) \cap J^-(\mathscr{I}^+) \setminus S$, Region 1. We call Region 2 the interior black hole region to the past of the shell, not including the shell, $J^-(S) \cap \left(J^-(\mathscr{I}^+)\right)^{\mathrm{c}} \setminus S$. By Region 3 we denote the exterior black hole region in the causal future of the shell, $J^+(S) \cap J^-(\mathscr{I}^+)$. We call Region 4 the interior black hole region in the causal future of the shell, $J^+(S) \cap \left(J^-(\mathscr{I}^+)\right)^{\mathrm{c}}$.

The $(1+1)$-dimensional version of the spacetime that we have just described is obtained by suppressing the angular coordinates. In the sequel, we find the mode functions for a Klein-Gordon field propagating in a $1+1$ collapsing null shell spacetime, with the aim of finding the Wightman function of the quantum theory.

\subsection{Construction of the two-point function}

We consider the real Klein-Gordon algebra $\mathscr{A}(M_{\rm V})$ for a massless, conformally coupled Klein-Gordon field in the null shell spacetime, and seek to construct a state $\omega: \mathscr{A}(M_{\rm V}) \to \mathbb{C}$ for the theory. We shall do this directly by obtaining the classical mode solutions. By doing this we in turn commit to a representation of the field on a Hilbert space.

We consider classical solutions to $\Box \varphi = 0$, where $\varphi: M_{\rm V} \to \mathbb{R}$, as incoming modes from infinity, which are of positive frequency with respect to $v$, appropriately defined in regions 1 and 3,  which in addition satisfy a Dirichlet boundary conditions at $r = 0$ in regions 1 and 2, so as to mimick the spherically-symmetric sector of the wave equation in $3+1$ dimensions. In regions 1 and 2, the mode solutions are
\begin{equation}
\Phi^{{1,2}}_\omega = -\frac{\ii}{(4 \pi \omega)^{1/2}} \left(\ee^{-\ii \omega u} -  \ee^{-\ii \omega v}\right),
\end{equation}
where $u = t - r$ in Regions 1 and 2 and $\omega$ is the mode frequency, while in Regions 3 and 4 one has
\begin{equation}
\Phi_\Omega^{{3,4}} = -\frac{\ii \ee^{\ii \Theta_\Omega}}{(4 \pi \Omega)^{1/2}} \left(G(U) -  \ee^{-\ii \Omega v}\right),
\end{equation}
where $\Omega$ is the mode frequency, $\Theta_\Omega$ is a phase and $G(U)$ is an as-of-yet undetermined function of the Kruskal $U$-coordinate. 

In order to determine the function $G(U)$, one can suppose that the phase $\Theta_\Omega = 0$ and that $\Omega = \omega$, and match the mode functions along the shell at $v = 0$. One then obtains the relation
\begin{equation}
G(U) = \ee^{\ii 2 \omega r} = \exp \left[ \ii \, 4 \omega M \left( 1 + W\left(-U/ \ee \right) \right) \right],
\end{equation}
where $W$ is the (strictly increasing) principal branch of the Lambert $W$-function \cite{NIST}, defined on the domain $(-1/\ee, \infty)$.\footnote{Recall that $-\infty < U < 1$ in the region of interest.} Thus, we have that
\begin{equation}
\Phi_\omega^{3,4} = -\frac{\ii}{(4 \pi \omega)^{1/2}} \left(\ee^{\ii \, 4 \omega M \left( 1 + W\left(-U/ \ee \right) \right)} -  \ee^{-\ii \omega v}\right).
\label{Phi34}
\end{equation}

The Wightman function can now be written down explicitly as a mode sum by standard methods, see e.g. \cite{Birrell:1982ix}. We are interested in obtaining the Wightman function in Regions 3 and 4. Define the function $\bar{u}(U) = - 4  M \left( 1 + W\left(-U/ \ee \right) \right)$. Then, the Wightman function defines a state by $\mathcal{W} \left(\x, \x'\right) := \langle \Omega | \widehat{\Phi}(\x) \widehat{\Phi}(\x') \Omega \rangle$, and it is given by
\begin{equation}
\mathcal{W}_\epsilon \left(\x, \x'\right) = -\frac{1}{4 \pi} \ln \left[\frac{\left(\bar{u} - \bar{u}' - \ii \epsilon\right)\left(v-v'-\ii \epsilon \right)}{\left(v - \bar{u}' - \ii \epsilon \right) \left( \bar{u} - v' - \ii \epsilon \right)} \right].
\label{W}
\end{equation}

It can be seen that the action of the isometries generated by the Killing vector $\xi = (4M)^{-1} (-U \partial_U + V \partial_V)$ in regions 3 and 4 does not leave the state \eqref{W} invariant.

%=====================================================================================================
% STRESS-ENERGY
%=====================================================================================================
\section{Comparison of the stress-energy tensors}
\label{sec:Stress-Energy}

In this section, we wish to compare the renormalised stress-energy tensor in the Unruh state, with state vector $|\Omega_{\rm U} \rangle$, and in the state that we have constructed in the collapsing null shell spacetime, with state vector $|\Omega \rangle$. This comparison is possible because, while the states are defined in different spacetimes, the union of Regions 3 and 4 of the collapsing null spacetime is isometric to the region of Kruskal spacetime delimited by $V \geq 1$, which is included in the union of the Kruskal future horizon, Regions I and II. We shall make this statement more precise below. First, we use the conformal techniques of Davies, Fulling and Unruh \cite{Birrell:1982ix, Davies:1976ei} to obtain the stress-energy tensors in the states $|\Omega_{\rm U} \rangle$ and $|\Omega \rangle$. The key point is that $1+1$ conformally flat spacetimes are conformal to (a region or all of) Minkowski spacetime.

Let $g$ and $\tilde{g}$ be two conformally related metrics, $g_{ab} = \Omega^2 \tilde{g}_{ab}$, where $\Omega^2 > 0$ is a conformal factor. In $1+1$ dimensions this relation takes the simple form $g = -\Omega^2 \dd u \dd v$ using appropriate coordinates. Then, we have that the renormalised stress-energy tensors are related by
\begin{align}
\langle \Omega | T_{ab}^{\rm ren} \Omega \rangle & = \langle \tilde{\Omega} | T_{ab}^{\rm ren} \tilde{\Omega} \rangle + \Theta_{ab} - (1/48\pi) {\rm R} \, g_{ab},
\label{Tab-ren-conf}
\end{align}
with $\Omega_{ab}$ locally defined as
\begin{subequations}
\label{Theta-ren-conf}
\begin{align}
\Theta_{uu} & = -(1/12\pi) \Omega \partial_u^2 \Omega^{-1}, \\
\Theta_{vv} & = -(1/12\pi) \Omega \partial_v^2 \Omega^{-1}, \\
\Theta_{uv} & = \Theta_{vu} = 0,
\end{align}
\end{subequations}
where $|\Omega \rangle$ and $|\tilde{\Omega}\rangle$ are conformally related state vectors defined on the spacetimes $(M,g)$ and $(\tilde{M}, \tilde{g})$ respectively, and ${\rm R}$ is the Ricci scalar of the spacetime $(M,g)$.

\subsection{Stress-energy tensor in the Unruh state}
\label{subsec:TabUnruh}

The Wightman function of the Unruh state is constructed from mode functions $\Phi_\omega = - \frac{\ii}{(4 \pi \omega)^{1/2}} \left(\ee^{-\ii \omega \bar{U}} - \ee^{-\ii \omega v} \right)$, where the first term on the right hand side is a right-moving mode, while the second is a left mover. Here $\bar{U} = 4M U$ has dimensions of length supplied by the inverse surface gravity at the horizon, a natural length scale that renders the exponent dimensionless. Let us write the Schwarzschild metric as
$g_{\rm S} = -\Omega^2 \dd \bar{U} \dd v$ with
\begin{equation}
\Omega^2(\bar{U},v) = \frac{2 M \ee^{-r/(2M)+v/(4M)}}{r} = -\frac{\left(1-2M/r\right)}{U},
\end{equation}
where $r$ is a function of $U$ (hence of $\bar{U}$) and $v$, as explained above, and the right-hand side is understood in a limiting sense as $r \to 2M$. This choice is made such that the two null coordinates in question are in each case the two null coordinates that define the positive frequency notion of the relevant mode functions. This allows us to say that the associated state in the associated flat spacetime is the Minkowski vacuum therein and has vanishing stress-energy. The Ricci scalar in $1+1$ Schwarzschild is ${\rm R} = 4M/r^3$, therefore the application of formula \eqref{Tab-ren-conf}, taking $\bar{U}$ and $v$ as the null coordinates, with respect to the Minkowski vacuum, $|\tilde{\Omega} \rangle = |\Omega_{\rm M}\rangle$ in Minkowski spacetime with the metric suitably written as $\tilde{g} = g_{\rm M} = -d\bar{U} dv$ yields
%\begin{subequations}
%\begin{align}
%\langle \Omega_{\rm U} | T_{ab}^{\rm ren} \Omega_{\rm U} \rangle & = \langle \Omega_{\rm U} | T_{\bar{U} \, \bar{U}}^{\rm ren} \Omega_{\rm U} \rangle \dd \bar{U}_a \dd \bar{U}_b + 2 \langle \Omega_{\rm U} | T_{\bar{U} \, v}^{\rm ren} \Omega_{\rm U} \rangle \dd \bar{U}_a \dd v_b + \langle \Omega_{\rm U} | T_{vv}^{\rm ren} \Omega_{\rm U} \rangle \dd v_a \dd v_b, \\
%\langle \Omega_{\rm U} | T_{\bar{U} \, \bar{U}}^{\rm ren} \Omega_{\rm U} \rangle & =  \frac{(2M-r)\left(4M^2 + r U + 2M(-3r+U)\right)}{96 \pi M U^2 r^4}, \\
%\langle \Omega_{\rm U} | T_{vv}^{\rm ren} \Omega_{\rm U} \rangle & =  - \frac{-192 M^4 + 128 M^3 r - 16 M^2 r^2 + r^4}{768 M^2 \pi r^4}, \\
%\langle \Omega_{\rm U} | T_{\bar{U} \, v}^{\rm ren} \Omega_{\rm U} \rangle &  = \langle \Omega | T_{v \, \bar{U}}^{\rm ren} \Omega \rangle =  -\frac{ \ee^{-r/(2M)} \ee^{v/(4M)}(1-2M/r)}{24 \pi r^3}.
%\end{align}
%\label{TabUnruh}
%\end{subequations}
%
\begin{subequations}
\begin{align}
\langle \Omega_{\rm U} | T_{ab}^{\rm ren} \Omega_{\rm U} \rangle & = \langle \Omega_{\rm U} | T_{\bar{U} \, \bar{U}}^{\rm ren} \Omega_{\rm U} \rangle \dd \bar{U}_a \dd \bar{U}_b + 2 \langle \Omega_{\rm U} | T_{\bar{U} \, v}^{\rm ren} \Omega_{\rm U} \rangle \dd \bar{U}_a \dd v_b + \langle \Omega_{\rm U} | T_{vv}^{\rm ren} \Omega_{\rm U} \rangle \dd v_a \dd v_b, \\
\langle \Omega_{\rm U} | T_{\bar{U} \, \bar{U}}^{\rm ren} \Omega_{\rm U} \rangle & =  \frac{(1-2M/r)^2}{48 \pi \bar{U}^2 r^2 }\left( 4 M r +r^2+ 12 M^2 \right), \\
\langle \Omega_{\rm U} | T_{vv}^{\rm ren} \Omega_{\rm U} \rangle & = \frac{M (3 M-2 r)}{48 \pi  r^4}, \\
\langle \Omega_{\rm U} | T_{\bar{U} \, v}^{\rm ren} \Omega_{\rm U} \rangle &  = \langle \Omega | T_{v \, \bar{U}}^{\rm ren} \Omega \rangle =  -\frac{M(1-2M/r)}{24 \pi U r^3}.
\end{align}
\label{TabUnruh}
\end{subequations}

\subsection{Stress-energy tensor in the collapsing null shell spacetime}

We are interested in obtaining the renormalised stress-energy tensor in Regions 3 and 4, where the Wightman function of the state that we have constructed is built out of the left and right-moving mode functions that appear in eq. \eqref{Phi34}, namely $\Phi_\omega^{3,4} = - \frac{\ii}{(4 \pi \omega)^{1/2}} \left(\ee^{-\ii \omega \bar{u}} - \ee^{-\ii \omega v} \right)$, where $\bar{u}$ can be related to the Kruskal $U$ coordinate by $\bar{u} = - 4  M \left( 1 + W\left(-U/ \ee \right) \right)$. We choose to write the metric $g_{\rm V}$ in Regions 3 and 4 as $g_{\rm V} = -\Omega^2 \dd \bar{u} \dd v$ with
\begin{equation}
\Omega^2(\bar{u},v) = -\frac{2M \ee^{-r/2M} \ee^{\kappa v}U \bar{u}}{(\bar{u}+4M) r} = \frac{\uu (1-2M/r)}{\uu + 4M},
\end{equation}
where $r$ is now viewed as a function of $\bar{u}$ and $v$, and $U$ as a function of $\uu$. The choice is made such that the stress-energy tensor in the conformally related spacetime vanishes by the positive frequency properties of the mode functions (cf. Sec. \ref{subsec:TabUnruh}).

The application of formula \eqref{Tab-ren-conf} yields
\begin{subequations}
\begin{align}
\langle \Omega | T_{ab}^{\rm ren} \Omega \rangle & = \langle \Omega | T_{\uu \uu}^{\rm ren} \Omega \rangle \dd {\uu}_a \dd {\uu}_b + 2 \langle \Omega | T_{\uu v}^{\rm ren} \Omega \rangle \dd {\uu}_a \dd v_b + \langle \Omega | T_{vv}^{\rm ren} \Omega \rangle \dd v_a \dd v_b, \\
\langle \Omega | T_{\uu \, \uu}^{\rm ren} \Omega \rangle & =  \frac{M \left(-16 (3 M+\uu) r^4 -2 \uu^4 r + 3 M \uu^4\right)}{48 \pi  \uu^2 (4 M+\uu)^2 r^4}, \\
\langle \Omega | T_{vv}^{\rm ren} \Omega \rangle & =  \frac{M (3 M-2 r)}{48 \pi  r^4}, \\
\langle \Omega | T_{\uu v}^{\rm ren} \Omega \rangle &  = \langle \Omega | T_{v \uu}^{\rm ren} \Omega \rangle = -\frac{M \uu(1-2M/r)}{12 \pi  (4 M+\uu) r^3}.
\end{align}
\label{TabShell}
\end{subequations}

\subsection{Comparison of the stress-energy tensors}
\label{subsec:Comp}

We wish to compare the stress-energy tensors as defined in the Unruh state and in the state constructed in the collapsing null shell spacetime. While the two states are defined on different spacetimes, there exist a region of the Kruskal spacetime, $(M_{\rm K}, g_{\rm K})$, that is isometric to the union of Regions 3 and 4 in the collapsing null shell spacetime $(M_{\rm V}, g_{\rm V})$. The key point is to view this region as a spacetime on its own right, $(M, g)$, and to define an algebra of observables and states in this ``late-time" spacetime. To make this statement precise, it is useful to take an algebraic approach based on \cite{Brunetti:2001dx}, motivated in turn by the ideas in \cite{Kay:1978zr}. See also \cite{Fewster:2006kt, Fewster:2006iy} in for a similar strategy in the context of quantum energy inequalities. 

\begin{figure}
  \centering
  \includegraphics[width=0.65\linewidth]{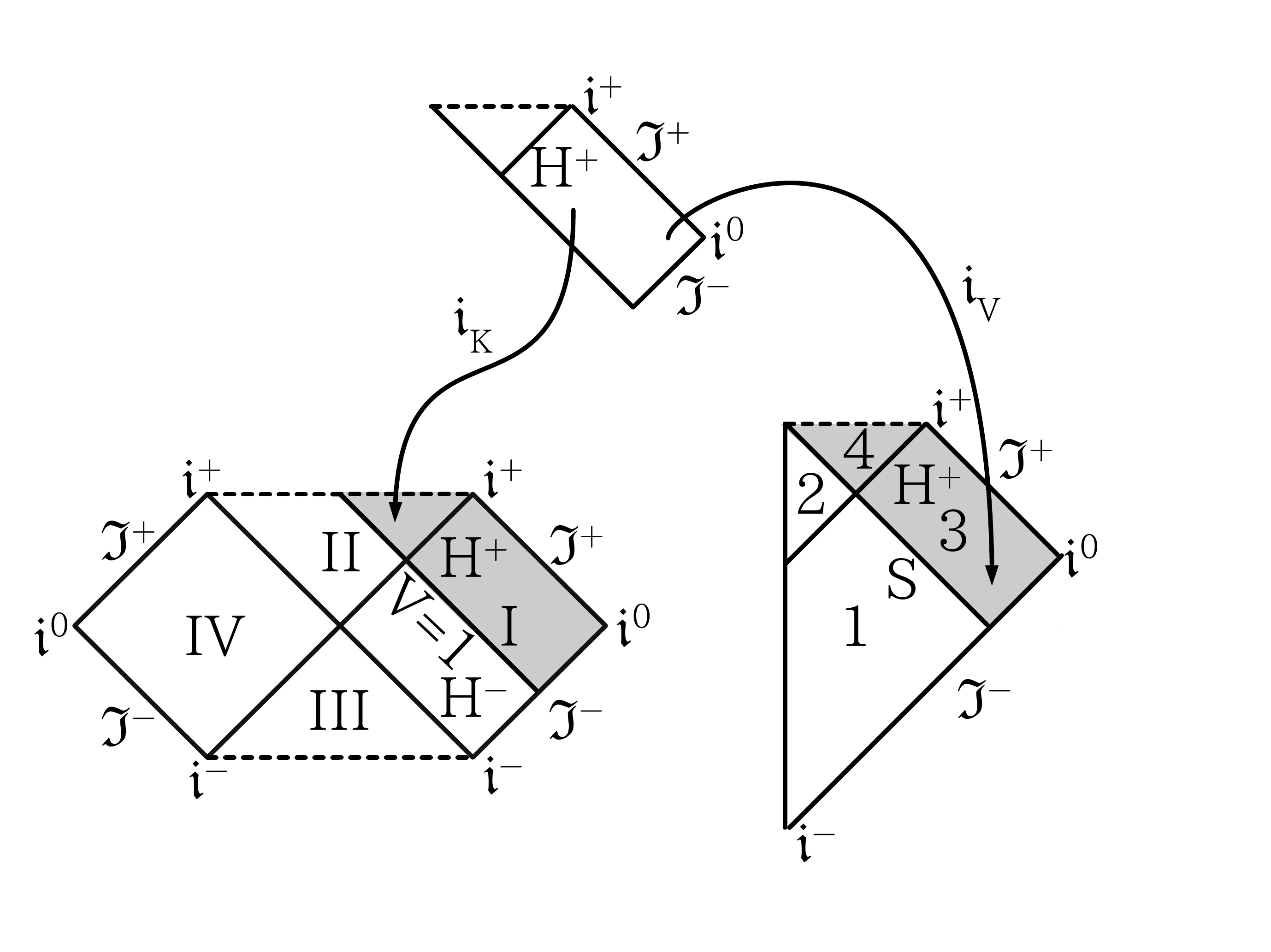}
  \caption{Isometric embeddings of the ``late-time" spacetime into the Kruskal spacetime and the collapsing null shell spacetime.}
  \label{Fig:Embed}
\end{figure}

The discussion that follows is necessarily abstract, and a reader interested in concrete results might proceed to Sec. \ref{subsec:NearHorizon}, considering that the punchline of the ensuing argument is that comparisons of expectation values of local observables (e.g. the stress-energy tensor) confined to the shaded regions in Fig. \ref{Fig:Embed} can be made in a precise sense.

Let $(M, g)$, which we denote the ``late-time" spacetime, be defined as the submanifold of $(M_{\rm V}, g_{\rm V})$ covered by Regions 3 and 4 in the collapsing null shell spacetime, with $M = J^+(S)$ (see Fig. \ref{Fig:Shell}) and $g$ the induced metric from $g_v$ on $M$, and let us call $i_{\rm V}: (M,g) \to (M_{\rm V}, g_{\rm V})$ the isometric embedding of $(M, g)$ into $(M_{\rm V}, g_{\rm V})$. There exists also an isometric embedding into Kruskal spacetime $i_{\rm K}: (M,g) \to (M_{\rm K}, g_{\rm K})$. See Fig. \ref{Fig:Embed}.
Associated with the ``late-time" spacetime is an algebra of observables $\mathscr{A}(M)$, whose elements are defined from algebra-valued distributions, mapping test functions, $f \in C_0^\infty(M)$, to algebra elements. Further, associated to the isometric embeddings, $i_{\rm V}$ and $i_{\rm K}$, are $^*$-preserving, unit-preserving homomorphisms, which are also injective (hence monomorphisms), $\mathscr{A}(i_{\rm V}):\mathscr{A}(M) \to \mathscr{A}(M_{\rm V}) $ and $\mathscr{A}(i_{\rm K}): \mathscr{A}(M) \to \mathscr{A}(M_{\rm K})$ respectively.\footnote{Here, we are considering that the algebra of observables has been suitably enlarged so as to contain the stress-energy tensor.} It follows from the homomorphism property that the kernels, $\ker[\mathscr{A}(i_{\rm V})]$ and $\ker[\mathscr{A}(i_{\rm K})]$ respectively, must be the ideal of $\mathscr{A}(M)$, but if $\mathscr{A}(M)$ is simple (as is our case), the ideal must be trivial or the whole algebra. In the latter case, the homomorphism cannot be unit-preserving, hence the kernels are trivial, which implies that $\mathscr{A}(i_{\rm V})$ and $\mathscr{A}(i_{\rm K})$ are invertibles, and hence $\mathscr{A}(M)$ is $^*$-isomorphic with the subalgebras ${\rm Im}[\mathscr{A}(i_{\rm V})]$ and ${\rm Im}[\mathscr{A}(i_{\rm K})]$.

To end the argument, we can pullback the state defined in the collapsing null shell spacetime, $\omega: \mathscr{A}(M_{\rm V}) \to \mathbb{C}$, and the Unruh state defined in the physical region of Kruskal (see Sec. \ref{sec:KG}), $\omega_{\rm U}: \mathscr{A}(M_{\rm S} \subset M_{\rm K}) \to \mathbb{C}$, to states in the ``late-time" spacetime, $\omega_{\rm V} = \mathscr{A}(i_{\rm V})^*\omega$ and $\omega_{\rm K} = \mathscr{A}(i_{\rm K})^*\omega_{\rm U}$ respectively.

We can therefore compare the expectation values of the renormalised stress-energy tensor with respect to the states $\omega_{\rm V}$ and $\omega_{\rm K}$ in the late-time spacetime. For simplicity in the book-keeping of notation, we henceforth continue to refer to $\omega_{\rm V}$ as $\omega$ (with state vector $|\Omega \rangle$) and to $\omega_{\rm K}$ as $\omega_{\rm U}$ (with state vector $|\Omega_{\rm U} \rangle$) in the ``late-time" spacetime.

\subsubsection{Near-horizon behaviour}
\label{subsec:NearHorizon}

We can readily compare the behaviour of the stress-energy tensor in the near-horizon region. For both the Unruh state and for the state produced out of the shell collapse we have that\footnote{For the purposes of our comparisons, we perform coordinate transformations on eq. \eqref{TabUnruh} and \eqref{TabShell} to bring the stress-energy tensors to an Eddington-Finkelstein coordinate basis in our region of interest.}
\begin{subequations}
\begin{align}
\langle \Omega | T_{v \, v}^{\rm ren} \Omega \rangle = \langle \Omega_{\rm U} | T_{v \, v}^{\rm ren} \Omega_{\rm U} \rangle & = -\frac{1}{768 \pi  M^2} +\frac{(r-2 M)^2}{512 \pi  M^4}-\frac{5 (r-2 M)^3}{1536 \pi  M^5} + O\left((r-2 M)^4\right), \label{TvvHorizon} \\
\langle \Omega | T_{u \, v}^{\rm ren} \Omega \rangle = \langle \Omega_{\rm U} | T_{v \, u}^{\rm ren} \Omega_{\rm U} \rangle &  = \frac{r-2 M}{384 \pi  M^3}-\frac{(r-2 M)^2}{192 \pi  M^4}+\frac{5(r-2 M)^3}{768 \pi  M^5}+O\left((r-2 M)^4\right). \label{TuvHorizon}
\end{align}
\label{Tvv-uvHorizon}
\end{subequations}

On the other hand, for the Unruh state,
\begin{align}
\langle \Omega_{\rm U} | T_{u \,u}^{\rm ren} \Omega_{\rm U} \rangle & =  \frac{(r-2 M)^2}{512 \pi  M^4 }-\frac{5 (r-2 M)^3}{1536 \left(\pi  M^5\right)} + O\left((r-2 M)^4\right), \label{TuuUnruhHorizon}
\end{align}
while for the collapsing shell state we have
\begin{align}
\langle \Omega | T_{u \,u}^{\rm ren} \Omega \rangle & = \frac{(r-2 M)^2}{512 \pi  M^4} \left(1-\ee^{-\frac{v}{2 M}}\right)-\frac{(r-2 M)^3 \left(5+ 3 \ee^{-\frac{2 v}{4 M}} -8 \ee^{-\frac{3 v}{4 M}} \right)}{1536 \pi  M^5} \nonumber \\
& +O\left((r-2 M)^4\right). \label{TuuShellHorizon}
\end{align}

Thus, we can see from eq. \eqref{Tvv-uvHorizon}, \eqref{TuuUnruhHorizon} and \eqref{TuuShellHorizon} that the near horizon behaviour of the stress-energy tensor of a Klein-Gordon field during stellar collapse is captured very precisely by the Unruh state, with deviation of order $O\left((r-2M)^2\right)$. In particular, the flux of negative energy that gives rise to black hole radiation, eq. \eqref{TvvHorizon}, is the dominant contribution of the stress-energy tensor in this regime.

\subsubsection{Near-future-infinity behaviour}

Near the future null infinity, at fixed $\uu$ and as $r \to \infty$ (or $v \to \infty$), we have on the one hand,
\begin{subequations}
\begin{align}
\langle \Omega | T_{v \, v}^{\rm ren} \Omega \rangle = \langle \Omega_{\rm U} | T_{v \, v}^{\rm ren} \Omega_{\rm U} \rangle & = -\frac{M}{24 \pi  r^3}+\frac{M^2}{16 \pi  r^4}+O\left(r^{-5}\right) , \label{TvvInfty} \\
\langle \Omega | T_{u \, v}^{\rm ren} \Omega \rangle = \langle \Omega_{\rm U} | T_{u \, v}^{\rm ren} \Omega_{\rm U} \rangle & = \frac{M}{24 \pi  r^3}-\frac{M^2}{12 \pi  r^4}+O\left(r^{-5}\right). \label{TuvInfty}
\end{align}
\end{subequations}
\label{Tvv-uvInfty}

On the other hand, for the Unruh state
\begin{align}
\langle \Omega_{\rm U} | T_{u \, u}^{\rm ren} \Omega_{\rm U} \rangle & = \frac{1}{768 \pi  M^2}-\frac{M}{24 \pi  r^3}+\frac{M^2}{16 \pi  r^4}+O\left(r^{-5}\right),
\label{TuuUnruhInfty}
\end{align}
while in the collapsing shell spacetime
\begin{align}
\langle \Omega | T_{u \, u}^{\rm ren} \Omega \rangle & = \frac{-3 M^2-M \uu}{3 \pi  \uu^4}-\frac{M}{24 \pi  r^3}+\frac{M^2}{16 \pi  r^4}+O\left(r^{-5}\right).
\label{TuuShellInfty}
\end{align}

From eq. \eqref{TuuShellInfty} it is clear that the radiation output to infinity is positive, but has a richer form compared to the output radiated in the Unruh state. We also note that as one approaches the future timelike infinity, i.e., as $\uu \to -4M$ along the future null infinity, the leading terms of eq. \eqref{TuuUnruhInfty} and \eqref{TuuShellInfty} coincide, $\langle \Omega | T_{u \, u}^{\rm ren} \Omega \rangle|_{\mathscr{I}^+} - \langle \Omega_{\rm U} | T_{u \, u}^{\rm ren} \Omega_{\rm U} \rangle|_{\mathscr{I}^+} = O\left((\uu + 4M)^2\right)$. Moreover, the leading term of eq. \eqref{TuuShellInfty} has non-decreasing derivative with respect to $\uu$, and therefore $\partial_u \langle \Omega | T_{u \, u}^{\rm ren} \Omega \rangle  \geq 0$ for $\uu \in (-\infty, -4M]$ indicating that at late $u$-time the output of radiation at $\mathscr{I}^+$ increases. Hence, we find that 
\begin{align}
0 \leq  \langle \Omega | T_{u \, u}^{\rm ren} \Omega \rangle|_{\mathscr{I}^+} \leq 1/\left(768 \pi M^2\right).
\label{TuuBoundInfty}
\end{align}

The upper bound in eq. \eqref{TuuBoundInfty} is attained exponentially fast as $u \to \infty$. This can be verified by analysing the asymptotic behaviour (cf. eq. \eqref{TuuShellInfty}) of 
\begin{equation}
F(u) = \frac{1}{768 \pi M^2} + \frac{3 M^2+M \uu}{3 \pi  \uu^4} = \frac{1}{768 \pi M^2} \left(1 - \frac{1-4 W\left(\ee^{-u/(4M)-1} \right)}{\left(1+ W\left(\ee^{-u/(4M)-1} \right) \right)^4} \right),
\end{equation}
with $(1- 4 W(z))/(1+W(z))^4 = 1-8 z + O\left(z^2\right)$.

\begin{figure}
  \begin{subfigure}[b]{0.45\textwidth}
    \includegraphics[width=\textwidth]{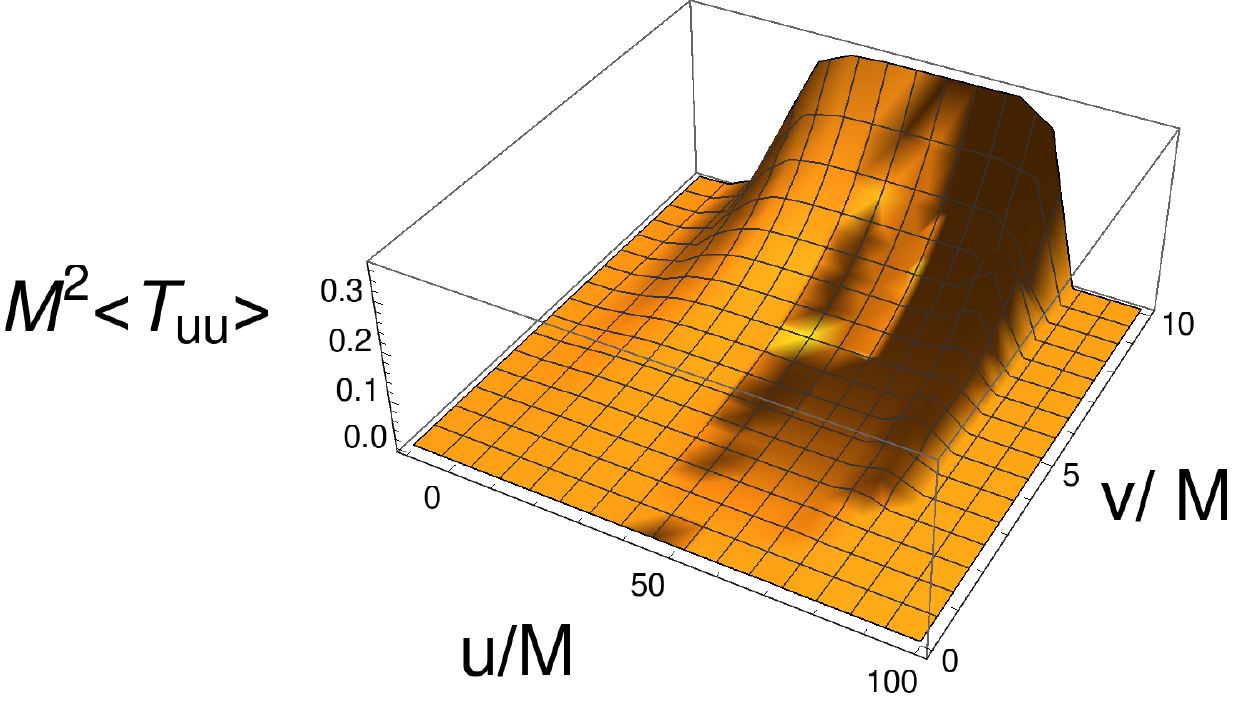}
    \caption{$\langle \Omega | T_{u \, u}^{\rm ren} \Omega \rangle$}
    \label{Fig:TuuShell}
  \end{subfigure}
  \begin{subfigure}[b]{0.45\textwidth}
    \includegraphics[width=\textwidth]{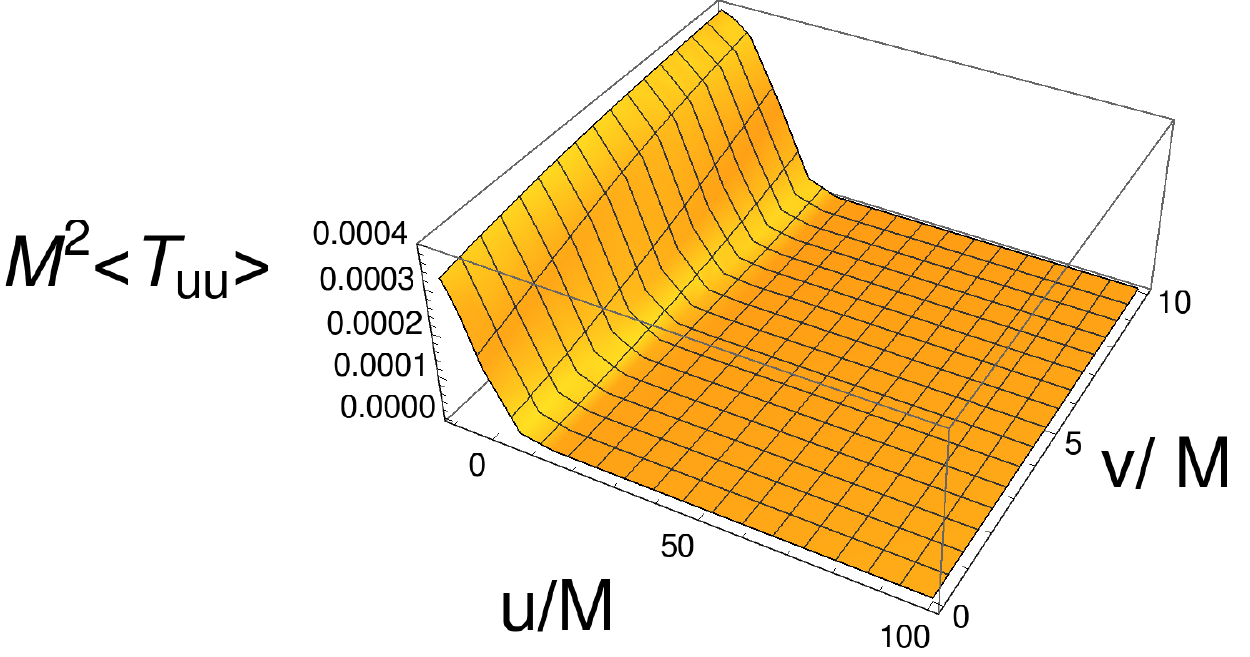}
    \caption{$\langle \Omega_{\rm U} | T_{u \, u}^{\rm ren} \Omega_{\rm U} \rangle$}
    \label{Fig:TuuUnruh}
  \end{subfigure}
  \caption{Sampling of the values taken in the ``late-time" spacetime (see Fig. \ref{Fig:Embed}) by $T_{uu}^{\rm ren}$, the renormalised stress-energy tensor, in (a) the state produced in the collapsing shell spacetime, $|\Omega\rangle$, and (b) the Unruh state, $|\Omega_{\rm U}\rangle$. Notice that as $u \to \infty$, near the horizon at $r = 2M$, the values of the expectation values coincide.}
  \label{Fig:Tuu}
\end{figure}

For the convenience of the reader, samplings of the values taken by $\langle \Omega | T_{u \, u}^{\rm ren} \Omega \rangle$ and $\langle \Omega_{\rm U} | T_{u \, u}^{\rm ren} \Omega_{\rm U} \rangle$ are plotted in Fig. \ref{Fig:Tuu}.

%=====================================================================================================
% DETECTORS
%=====================================================================================================
\section{The experience of late-time observers}
\label{sec:Detectors}

We now analyse the experience of an observer carrying a detector in the collapsing null shell spacetime, and analyse the emergence of radiation at late times for an orbit of constant radial coordinate $ r > 2M$.

The formula for the response of an Unruh-DeWitt detector coupled to a quantum scalar in $1+1$ dimensions with a derivative-coupling interaction \cite{Juarez-Aubry:2014jba}, $H_{\rm int}(\tau) = c \chi(\tau) \mu(\tau) \dot{\Phi}(\rm{x}(\tau))$, where $\tau$ is the proper time along the detector's worldline, $c$ is a coupling constant, $\chi$ is the detector's switching function (typically smooth of compact support), $\mu$ is the monopole moment of the detector, and $\dot{\Phi}(\tau)$ is the proper-time derivative of the pullback of the field to the detector worldline, is given by
\begin{align}
\mathcal{F}(\omega) = & - \frac{\omega}{2} \int_{-\infty}^{\infty} \! du \, [\chi(u)]^2 + \frac{1}{\pi} \int_0^\infty \! \frac{ds}{s^2} \, \int_{-\infty}^\infty \! du \, \chi(u) [\chi(u) - \chi(u-s)] \nonumber \\
& + 2 \int_{-\infty}^\infty \! du \, \int_0^\infty \chi(u) \chi(u-s) {\rm Re} \left[\ee^{-\ii \omega s} \mathcal{A}(u, u-s) + \frac{1}{2 \pi s^2} \right],
\label{Response}
\end{align}
with $\mathcal{A}(\tau, \tau') = \partial_\tau \partial_{\tau'} \mathcal{W}(\tau, \tau')$, where by $\mathcal{W}(\tau, \tau')$ we mean the pullback of the Wightman two-point bi-distribution to the worldline of the detector, and where the derivatives should be understood in a distributional sense. In the formula \eqref{Response}, the integrand is a {\it bona fide} function free of distributional singularities for Hadamard states. This is so because the subtraction of  $-(2 \pi s)^{-2}$ to $\mathcal{A}(u, u-s)$ takes care of the distributional singularities arising from the Hadamard expansion in the short-distance limit of the Wightman bi-distribution.

We use the derivative coupling because, while the two-point function for the scalar field in the collapsing-shell spacetime has no infrared ambiguities, the Unruh state two-point function is ambiguous, and we wish to compare the two responses on equal grounds.

Here, we consider that the detector is switched-on sharply at some time $\tau_0$ and that its rate is read at very late times, as $\tau \to \infty$ along the worldline. We choose for concreteness that the detector is switched on when the detector crosses the shell, at $v = 0$, but this choice is irrelevant for our late-time estimates.

In the sharp switching limit, the detector response diverges logarithmically as the switching time vanishes, but the rate, $\dot{\mathcal{F}}$ remains finite and is given by \cite{Juarez-Aubry:2014jba}
\begin{equation}
\dot{\mathcal{F}}(\omega, \Delta \tau) = -\frac{\omega}{2} + \frac{1}{\pi\Delta\tau} + 2 \int_0^{\Delta \tau} \! ds \, {\rm Re} \left[\ee^{-\ii \omega s} \mathcal{A}(\tau, \tau-s) + \frac{1}{2 \pi s^2} \right],
\label{SharpRate}
\end{equation}
where $\Delta \tau = \tau - \tau_0$ is the total interaction detector proper time between the detector and the field, such that the detector is switched on at time $\tau_0$ and read at time $\tau$.

For a fixed orbit, the shell crossing occurs at $v = 0$ and $r = R$ and we choose the crossing to occur at $\tau_0 = 0$. %Analogously, the crossing point is at $(t, r) = - R - 2M \ln(R/2M-1), R)$ or at $(U,V) = ([1-R/(2M)] \ee^{R/(2M)}, 1)$. 
In practice, we can regard the experience of these observers as restricted to the late-time spacetime, and compare the detector rate for a Klein-Gordon field in the Unruh state and in the collapse null shell spacetime state.

For the Unruh state, the late-time behaviour of the detector rate moving with fixed $r = R$ is known to capture the Hawking radiation coming from the right-movers (with the Hawking temperature weighted by an appropriate Tolman factor), while the left-moving modes contribute as if the state where the Minkowski state. This can be seen from eq. \eqref{SharpRate} in the limit $\tau \to \infty$ with the bi-distribution
\begin{align}
\mathcal{A}^{\rm U}_\epsilon \left(\tau, \tau'\right) = & -\frac{1}{4 \pi}  \left[ \frac{\dot{U} \,\dot{U}'}{(U-U'-\ii \epsilon)^2} +  \frac{\dot{v} \, \dot{v}'}{(v-v'-\ii \epsilon)^2} \right],
\label{AU}
\end{align}
obtained from the Unruh state, where the distributional character of $\mathcal{A}^{\rm U}$ is encoded in the $\epsilon \to 0^+$ limit of $\mathcal{A}_\epsilon^{\rm U}$,
\begin{align}
\mathcal{W}_\epsilon^{\rm U}({\rm x}, {\rm x}') = -\frac{1}{4 \pi }  \ln\left[(\epsilon + \ii (\bar{U}-\bar{U}'))(\epsilon + \ii(v - v'))   \right].
\label{WUamb}
\end{align}

Here, the definition of $\mathcal{W}_\epsilon^{\rm U}$ by the right-hand side of eq. \eqref{WUamb} is unique up to the addition of an ambiguous real-valued constant. Notice that $\mathcal{A}^{\rm U}_\epsilon$ can be seen as a function of the difference $\tau - \tau'$ along a stationary orbit (generated by $\xi = \partial_t$), $r = R > 2M$, due to the invariance, up to a constant, of $\mathcal{W}_\epsilon^{\rm U}$ under the action of the isometries generated by $\xi$. One finds that the rate at late times is \cite{Juarez-Aubry:2014jba},
\begin{equation}
\lim_{\Delta \tau \to \infty} \dot{\mathcal{F}}^{\rm U}(\omega, \Delta \tau) = -\frac{\omega}{2} \Theta(-\omega) + \frac{\omega}{2 \left(\ee^{\omega/ T_{\rm loc}} - 1\right)} + o(1),
\label{UnruhResponse}
\end{equation}
where $\Theta$ is the Heaviside step function, and with the local temperature defined as $T_{\rm loc} = (1-2M/R)^{-1/2} T_{\rm H}$, where $T_{\rm H} = 1/(8 \pi M)$ is the Hawking temperature.

For the collapsing shell state, we have that
\begin{align}
\mathcal{A}_\epsilon \left(\tau, \tau'\right) =  & -\frac{1}{4 \pi}  \left[ \frac{\dot{\bar{u}} \,\dot{\bar{u}}'}{(\bar{u}-\bar{u}'-\ii \epsilon)^2} +  \frac{\dot{v} \, \dot{v}'}{(v-v'-\ii \epsilon)^2} \right. \nonumber \\
& \left. - \frac{\dot{v} \, \dot{\bar{u}}'}{(v-\bar{u}'-\ii \epsilon)^2} - \frac{\dot{\bar{u}} \, \dot{v}'}{(\bar{u}- v' - \ii \epsilon)^2} \right],
\label{A}
\end{align}
and using formula \eqref{SharpRate} we can write the rate as a sum of several contributions,
\begin{subequations}
\label{Fshell}
\begin{align}
\dot{\mathcal{F}}(\omega, \tau) & = \dot{\mathcal{F}}_{\bar{u} \, \bar{u}}(\omega, \tau) + \dot{\mathcal{F}}_{v \, v}(\omega, \tau) + \dot{\mathcal{F}}_{v \, \bar{u}}(\omega, \tau) + \dot{\mathcal{F}}_{\bar{u} \, v}(\omega, \tau), \\
\dot{\mathcal{F}}_{\bar{u} \, \bar{u}}(\omega, \tau) & = \frac{1}{2 \pi} \int_0^{\Delta \tau} \! ds \, \left[-\cos(\omega s) \frac{\dot{\bar{u}}(\tau) \,\dot{\bar{u}}(\tau-s)}{[\bar{u}(\tau)-\bar{u}(\tau-s)]^2} + \frac{1}{s^2} \right], \label{Fuu} \\
\dot{\mathcal{F}}_{v \, v}(\omega, \tau) & = \frac{1}{2 \pi} \int_0^{\Delta \tau} \! ds \, \left[-\cos(\omega s) \frac{\dot{v}(\tau) \,\dot{v}(\tau-s)}{[v(\tau)-v(\tau-s)]^2} + \frac{1}{s^2} \right] - \frac{\omega}{2}, \\
\dot{\mathcal{F}}_{v \, \bar{u}}(\omega, \tau) & = \frac{1}{2 \pi}  \int_0^{\Delta \tau} \! ds \,\cos(\omega s) \frac{\dot{v}(\tau) \,\dot{\bar{u}}(\tau-s)}{[v(\tau)-\bar{u}(\tau-s)]^2}, \label{Fvu} \\
\dot{\mathcal{F}}_{\bar{u} \, v}(\omega, \tau) & = \frac{1}{2 \pi} \int_0^{\Delta \tau} \! ds \, \cos( \omega s) \frac{\dot{\bar{u}}(\tau) \,\dot{v}(\tau-s)}{[\bar{u}(\tau)-v(\tau-s)]^2}, \label{Fuv}
\end{align}
\end{subequations}
where each piece has been organised in such a way that each integrand is a singularity-free expression. 

As is the case in the Unruh state, the purely left-moving contribution contributes like the Minkowski state, $\dot{\mathcal{F}}_{v \, v}(\omega, \tau) = - (\omega/2) \Theta(- \omega)$. We show in Appendix \ref{app:A} that as $\tau \to \infty$, we have $\dot{\mathcal{F}}_{v \, \bar{u}}(\omega, \tau) = \dot{\mathcal{F}}_{\bar{u} \, v}(\omega) = o(1)$, while $\dot{\mathcal{F}}_{\bar{u} \, \bar{u}}(\omega, \tau)$ contributes as a Planckian spectrum at the expected temperature. Namely, eq. \eqref{Fshell} yields
\begin{equation}
\dot{\mathcal{F}}(\omega, \tau) = -\frac{\omega}{2} \Theta(-\omega) + \frac{\omega}{2 \left(\ee^{\omega/ T_{\rm loc}} - 1\right)} + o(1),
\label{ShellResponse}
\end{equation}
and we conclude that the late-time transition rate of the Unruh state is in excellent agreement with the late-time rate in the state of the collapsing shell spacetime.

\begin{figure}
  \begin{subfigure}[b]{0.3\textwidth}
    \includegraphics[width=\textwidth]{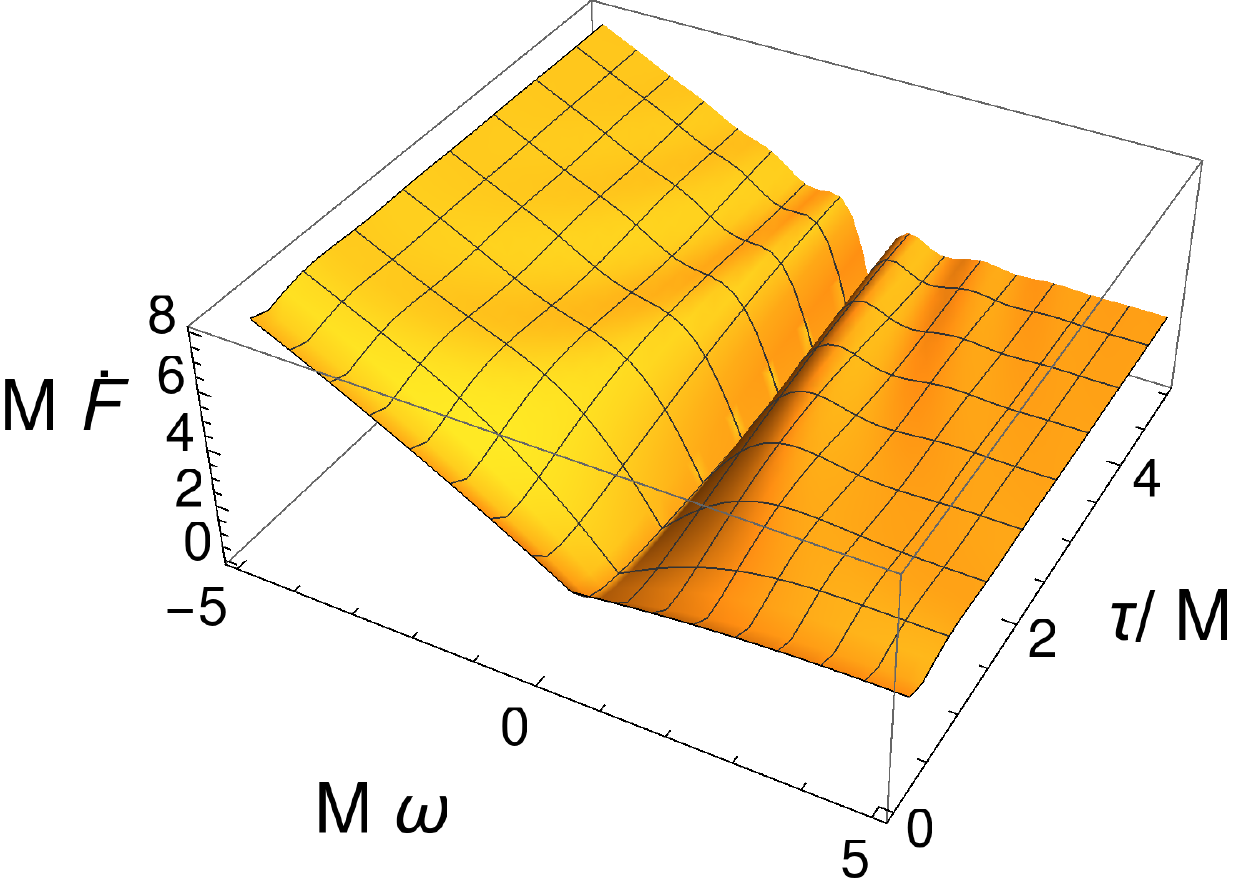}
    \caption{$\dot{\mathcal{F}}(\omega, \Delta \tau)$}
    \label{Fig:RateShell}
  \end{subfigure}
  \begin{subfigure}[b]{0.3\textwidth}
    \includegraphics[width=\textwidth]{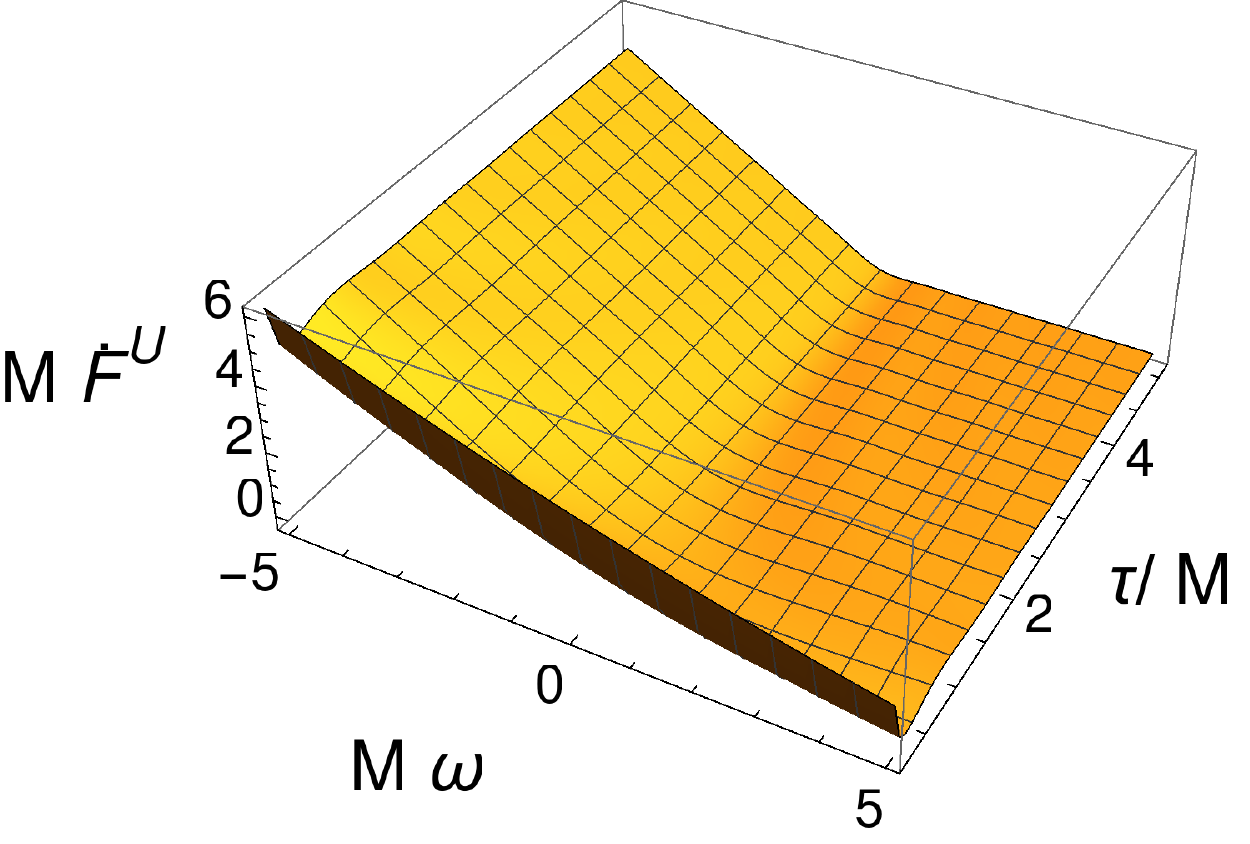}
    \caption{$\dot{\mathcal{F}}^{\rm U}(\omega, \Delta \tau)$}
    \label{Fig:RateUnruh}
  \end{subfigure}
  \begin{subfigure}[b]{0.3\textwidth}
    \includegraphics[width=\textwidth]{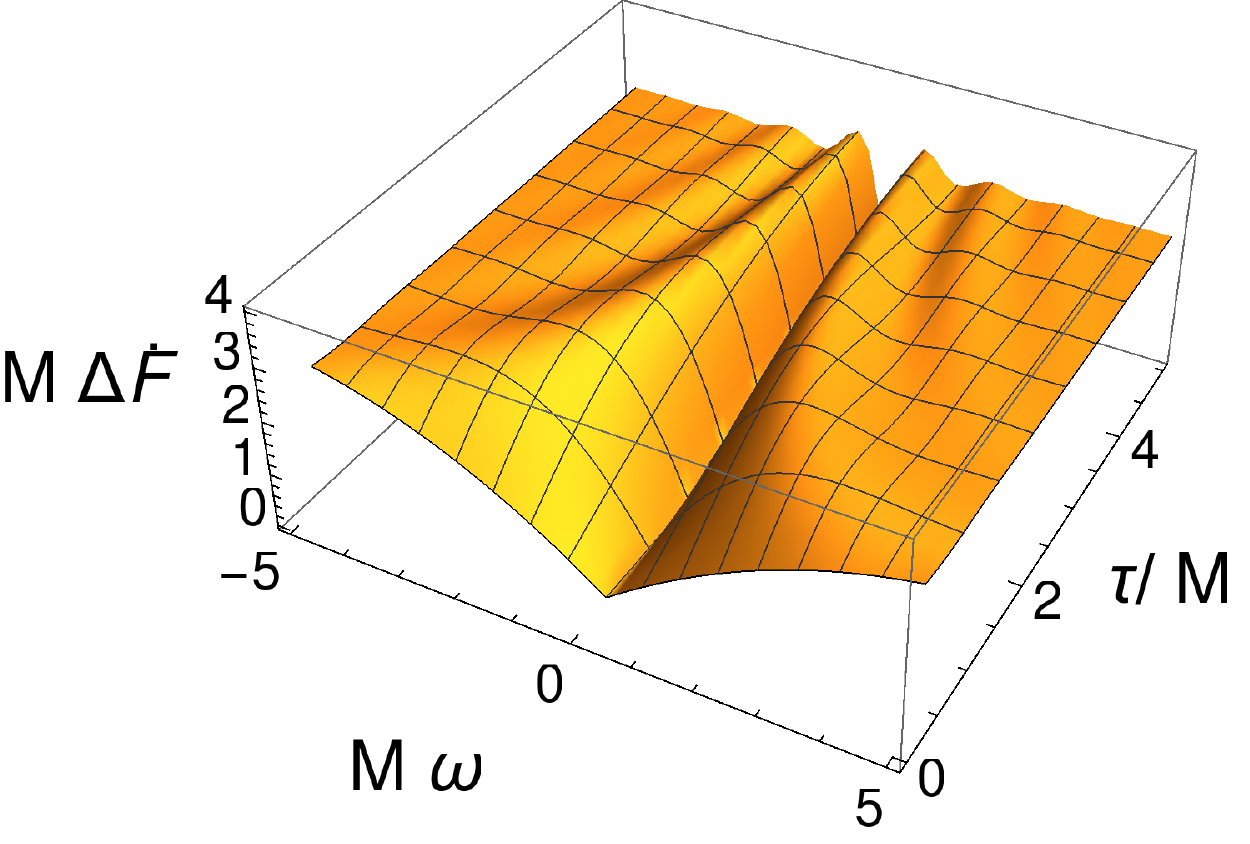}
    \caption{$\dot{\mathcal{F}}(\omega, \Delta \tau) - \dot{\mathcal{F}}^{\rm U}(\omega, \Delta \tau)$}
    \label{Fig:Difference}
  \end{subfigure}
  \caption{Comparison of the detector transition rates measured at finite times $\tau > \tau_0$ for a detector fixed at $r = 3M$ and switched on sharply at proper time $\tau_0 = 0$ at $(v, r) = (0, r)$. Fig. \ref{Fig:RateShell} shows the rate in the collapsing shell spacetime state. Fig. \ref{Fig:RateUnruh} shows the rate in the Unruh state. Fig. \ref{Fig:Difference} displays the difference between the two rates.}
  \label{Fig:Rates}
\end{figure}

The early time behaviour of the Unruh and the collapsing shell spacetime's states can be explored numerically. We show in Fig. \ref{Fig:Rates} the finite-time discrepancies between the transition rates of detectors in the two states, such that the detector is sharply switched on at proper time $\tau_0 = 0$ at the spacetime point $(v, r) = (0, R)$ and measured at some later finite time $\tau$. It can be seen numerically that, unlike in the case of the Unruh state, the detector rate is not steadily decreasing in the detector gap, $\omega$, in the collapsing shell scenario, and the onset of thermality takes place only at late times.

%=====================================================================================================
% CONCLUSIONS
%=====================================================================================================
\section{Conclusions}
\label{sec:conc}

In this work, we have asked and answered the question of how good the usual folklore that treats the Unruh state in Schwazschild spacetime as the state emerging from physical black hole formation is. We have done so by analysing the simple model in $1+1$ dimensions of a Vaidya spacetime consisting of a collapsing null shell that forms a black hole, in which the two-point function can be computed explicitly, and the stress-energy tensor obtained using conformal techniques.

Our findings are that the Unruh state provides an excellent estimate in the near-horizon region of the spacetime, and that the negative energy flux computed from the Unruh state matches very precisely a negative energy flux present at the horizon in the collapsing null shell spacetime. The behaviour of the radiation flux output at future null infinity is not completely captured by the Unruh state, but near future timelike infinity the outgoing flux in the collapsing shell spacetime is well characterised by the Unruh state. Moreover, we find that pointwise $0 \leq  \langle \Omega | T_{u \, u}^{\rm ren} \Omega \rangle|_{\mathscr{I}^+} \leq  \langle \Omega_{\rm U} | T_{u \, u}^{\rm ren} \Omega_{\rm U} \rangle|_{\mathscr{I}^+}$, i.e., at every point on $\mathscr{I}^+$ the outgoing flux of radiation in the Unruh state dominates the radiation output in the collapsing null shell spacetime, and the latter is strictly non-decreasing in the $u$-time at future null infinity ($v \to \infty$). The energy output at infinity predicted using calculations based on the Unruh state, $\langle \Omega_{\rm U} | T_{u \, u}^{\rm ren} \Omega_{\rm U} \rangle|_{\mathscr{I}^+} = 1/(768 \pi M^2)$ is however approached exponentially fast in $u$-time. Thus, back-reaction estimates of pre-Hawking radiation based on the Unruh state already over-estimate the radiation output, and this makes it unlikely, in our view, that pre-Hawking radiation can prevent black-hole formation.

We have also analysed the character of the radiation as perceived by an external local observer, moving on at fixed radial coordinate $r > 2M$ in the causal future of the shell, carrying a particle detector that couples to the derivative of the field (in order to avoid infrared ambiguities). We have found that at late times, near future timelike infinity, the right-moving modes yield a thermal spectrum at the local Hawking temperature, in agreement with the Unruh state calculations, and showing the onset of thermality at late times. The detection of particles at earlier times has been explored numerically, showing substantial deviations between the rates measured by detectors in the collapsing shell spacetime and those measured from the coupling to a field in the Unruh state.

Finally, our $(1+1)$-dimensional analysis is relevant in the $3+1$ setting. In the case of the stress-energy tensor, the arguments of \cite{Ring:2006fk} show how to extrapolate the lower-dimensional renormalised stress-energy tensor to estimate the leading behaviour of the $(3+1)$-dimensional object. For detectors, the derivative coupling detector in $1+1$ captures the ultraviolet behaviour of the Wightman function in the integrand of the response function of an Unruh-DeWitt (non-derivative coupling) detector in $3+1$ dimensions.

\vspace*{1cm}

%\noindent \textbf{Note added in proof of published JHEP version:} After the completion of this work we became aware of
%\cite{Kawai:2013mda, Kawai:2014afa, Kawai:2015uya, Kawai:2017txu}, which are related to \cite{Baccetti:2016lsb, Baccetti:2016dzy, Baccetti:2017ioi, Baccetti:2017oas}. We thank Yuki Yokokura for bringing these papers to our attention. 

%=====================================================================================================
% ACKNOWLEDGMENTS
%=====================================================================================================
\section*{Acknowledgments}

BAJ-A was supported by CONACYT project 101712, Mexico. JL was supported in part by Science and Technology Facilities Council (Theory Consolidated Grant ST/P000703/1). BAJ-A also acknowledges the hospitality of the University of Nottingham, where the early stages of this work were discussed, as well as the support of an International Mobility Award granted by the Red Tem\'atica de F\'isica de Altas Energ\'ias (Red FAE) of CONACYT, and the additional support of PAPIIT- UNAM grant IG100316, Mexico. BAJ-A thanks Christopher J. Fewster for a very helpful email exchange concerning algebraic aspects  of QFT appearing in Sec. \ref{subsec:Comp}. Finally, we thank Yuki Yokokura for bringing the work in \cite{Kawai:2013mda, Kawai:2014afa, Kawai:2015uya, Kawai:2017txu} to our attention.
\appendix
%=====================================================================================================
% APPENDIX
%=====================================================================================================
\section{Auxiliary results for detector rates}
\label{app:A}

In this appendix, we show how to compute the expressions given by eq. \eqref{Fuu}, \eqref{Fvu} and \eqref{Fuv} in the limit $\tau \to \infty$. The key point will be to analyse the integrands and to apply the convergence theorems that allow us to apply the limit inside the integral.

\subsection{$\dot{\mathcal{F}}_{\bar{u} \, \bar{u}}$}

Let us begin by analysing the integrand of the expression defining $\dot{\mathcal{F}}_{\bar{u} \, \bar{u}}$ in the right hand side of eq. \eqref{Fuu}. We begin by writing
\begin{align}
\frac{\dot{\bar{u}}(\tau) \,\dot{\bar{u}}(\tau-s)}{[\bar{u}(\tau)-\bar{u}(\tau-s)]^2} & = \frac{\dot{u}(\tau) \dot{u}(\tau-s)}{(-4M)^2} \frac{W(-U(\tau-s)/\ee)}{W(-U(\tau)/\ee)} \nonumber \\
& \times \frac{[(1+W(-U(\tau)/\ee))(1+W(-U(\tau-s)/\ee))]^{-1}}{(1-W(-U(\tau-s)/\ee)/W(-U(\tau)/\ee))^2},
\end{align}
and using the defining relation of the Lambert W-function, $W(z) = z \ee^{-W(z)}$, we can write along the orbit, $(u, r) = (t - R - 2M \ln(R/2M-1), R)$, with the $t$ coordinate satisfying $t - t' = (1-2M/R)^{-1/2} (\tau - \tau')$,
\begin{align}
& \frac{\dot{\bar{u}}(\tau) \,\dot{\bar{u}}(\tau-s)}{[\bar{u}(\tau)-\bar{u}(\tau-s)]^2} \nonumber \\
& = \frac{(4M)^{-2}(1-2M/R)^{-1}[(1+W(-U(\tau)/\ee))(1+W(-U(\tau-s)/\ee))]^{-1}}{4\sinh^2\{[-W(-U(\tau-s)/\ee) + W(-U(\tau)/\ee)+(4M)^{-1}(1-2M/R)^{-1/2} s]/2\}}.
\label{F1Monotone}
\end{align}

In this form, one can readily verify that right hand side of eq. \eqref{F1Monotone} is strictly non-decreasing when viewed as a function of $\tau$. This follows from standard properties of the Lambert $W$-function, which is positive and non-decreasing for positive argument, from where it follows that $W(-U(\tau)/\ee)>0$ and $\partial_\tau W(-U(\tau)/\ee) \leq 0$ and, hence, that the numerator,
\begin{subequations}
\begin{align}
N(\tau,s) & =  [(1+W(-U(\tau)/\ee))(1+W(-U(\tau-s)/\ee))]^{-1} > 0, \\
\partial_\tau N(\tau,s) & \geq 0,
\end{align}
\end{subequations}
is a positive, non-decreasing function of $\tau$. For the denominator, notice that the argument of the $\sinh^2$ is a non-increasing function of $\tau$, and hence the denominator is non-increasing. It follows that one can apply the monotone convergence theorem and write eq. \eqref{Fuu} as
\begin{equation}
\lim_{\tau \to \infty} \dot{\mathcal{F}}_{\bar{u} \, \bar{u}}(\omega, \tau) = \frac{1}{2 \pi} \int_0^{\infty} \! ds \, \left[-\cos(\omega s) \lim_{\tau \to \infty} \left( \frac{\dot{\bar{u}}(\tau) \,\dot{\bar{u}}(\tau-s)}{[\bar{u}(\tau)-\bar{u}(\tau-s)]^2} \right) + \frac{1}{s^2} \right],
\label{FuuMonotone}
\end{equation}
where
\begin{equation}
\lim_{\tau \to \infty} \frac{\dot{\bar{u}}(\tau) \,\dot{\bar{u}}(\tau-s)}{[\bar{u}(\tau)-\bar{u}(\tau-s)]^2} = \frac{(1-2M/R)^{-1}}{4(4M)^2 \sinh^2[s/(8M(1-2M/R)^{1/2})] } +  o(1).
\label{FuuMonotone2}
\end{equation}

But eq. \eqref{FuuMonotone} with the first term in the integrand replaced by \eqref{FuuMonotone2} can be handled as a stationary problem by the complex analytic tecniques appearing in \cite[Sec. 2]{Juarez-Aubry:2014jba} and put in the form of formula 3.985.1 in \cite{Gradshteyn}, see \cite[Sec. 3.2]{Juarez-Aubry:2014jba}. One obtains at late proper time
\begin{equation}
\dot{\mathcal{F}}_{\bar{u} \, \bar{u}}(\omega, \tau) = -\frac{\omega}{2} \Theta(-\omega) + \frac{\omega}{2 \left(\ee^{\omega/ T_{\rm loc}} - 1\right)} + o(1).
\end{equation}

\subsection{$\dot{\mathcal{F}}_{v \, \bar{u}}$}

The second term can be computed using a similar strategy. We have that the integrand of \eqref{Fvu} can be written as
\begin{align}
\frac{ \dot{v}(\tau) \, \dot{\bar{u}}(\tau-s) }{[v(\tau)-\bar{u}(\tau-s)]^2} & = -\frac{ \dot{u}(\tau) \dot{v}(\tau-s) W(-U(\tau-s)/\ee)/(4M) }{[1+W(-U(\tau-s)/\ee)][v(\tau) + 4M + 4M W(-U(\tau-s)/\ee)]^2}.
\label{FvuInt}
\end{align}

The right hand side of eq. \eqref{FvuInt} vanishes as $\tau \to \infty$. We now proceed to show by a monotone convergence argument that we can take the limit inside the integral and, hence, we show that the contribution of $\dot{\mathcal{F}}_{v \, \bar{u}} = o(1)$. It suffices to study the $\tau$ derivative of the integrand at fixed $s$ of the integrand \eqref{FvuInt}. Recall that $\partial_\tau W(-U(\tau-s)/\ee) \leq 0$. Hence, for
\begin{align}
& \partial_\tau \frac{W(-U(\tau-s)/\ee) }{[1+W(-U(\tau-s)/\ee)][v(\tau) + 4M + 4M W(-U(\tau-s)/\ee)]^2} \nonumber \\
& = \frac{\partial_\tau  W(-U(\tau-s)/\ee) }{[1+W(-U(\tau-s)/\ee)]^2[v(\tau) + 4M + 4M W(-U(\tau-s)/\ee)]^2} \nonumber \\
& \times \left[ 1 - 8M \frac{W(-U(\tau-s)/\ee)}{v(\tau) + 4M + 4M W(-U(\tau-s)/\ee)} \right] \nonumber \\
& -  \frac{2 W(-U(\tau-s)/\ee) \dot{v}(\tau) }{[1+W(-U(\tau-s)/\ee)][v(\tau) + 4M + 4M W(-U(\tau-s)/\ee)]^3},
\label{DerivativeFvuInt}
\end{align}
one can see that the second term is non-positive, while for the first term, the first factor is non-positive, with the second factor being non-negative for sufficiently large $\tau$. Namely, it is guaranteed that when $\tau$ is sufficiently large, such that
\begin{equation}
v(\tau) \geq \sup_{s \in [0, \tau]} 4M [W(-U(\tau-s)/\ee)-1],
\end{equation}
the derivative \eqref{DerivativeFvuInt} is non-positive. It then follows from a monotone convergence argument that in the appropriate limit $\dot{\mathcal{F}}_{v \, \bar{u}}(\omega, \tau)  = o(1)$.

\subsection{$\dot{\mathcal{F}}_{\bar{u} \, v}$}

Finally, let us calculate the contribution of $\dot{\mathcal{F}}_{\bar{u} \, v}$. The integrand vanishes as $\tau \to \infty$, and we can take the limit inside the integral by a dominated convergence argument, as follows:
\begin{align}
\left| \frac{ \cos( \omega s) \, \dot{\bar{u}}(\tau) \,\dot{v}(\tau-s)}{2 \pi [\bar{u}(\tau)-v(\tau-s)]^2} \right| & \leq  \frac{C_1 \, W(-U(\tau)/\ee) \dot{u}(\tau) \dot{v}(\tau-s)}{[1+W(-U(\tau)/\ee)] [-4M - 4M W(-U(\tau)/\ee) - v(\tau-s)]^2} \nonumber \\
& \leq  \frac{C_2 }{[-4M - 4M W(-U(\tau)/\ee) - v(\tau-s)]^2} \leq \frac{C_3}{[4M + v(\tau-s)]^2}.
\end{align}
where $C_1$, $C_2$ and $C_3$ are positive constants. The right hand side of the expression above is integrable because $v(\tau) = t(\tau) + R + 2M \ln(R/2M -1)$ is linear in $\tau$. Hence $\dot{\mathcal{F}}_{\bar{u} \, v} = o(1)$ as $\tau \to \infty$.

%=====================================================================================================
% BIBLIOGRAPHY
%=====================================================================================================

\end{document}